[2018 Summer/Fall semester URP] KAIST URP Program

# Individual Research Project Final Report

Research Participation Project

## 성층권과 지상국의 공중 플랫폼 간의 다운 링크 자유 공간 광학 링크의 정렬 성능에 관한 연구

## Research on the alignment performance of downlink free-space optic link between airborne platforms in the stratosphere and ground station

**School of Electrical Engineering (20150923)**

# For Submission

**성층권과 지상국의 공중 플랫폼 간의 다운 링크 자유 공간 광학 링크의 정렬 성능에 관한 연구**

**Research on the alignment performance of downlink free-space optic link between airborne platforms in the stratosphere and ground station**

**Research Period : June 25, 2018 ~ December 14, 2018**

**Advisory Professor: Prof. Hoon Kim (Signature)**

**Teaching Assistant: Mai Viet Vuong (Signature)**

**As a participant(s) of the KAIST URP program, I (We) have completed the above research and hereby submit the final report on the research.**

**Month, Date, 2018**

**Researcher Bereket Eshete (Signature)**

# Contents

# Abstract

Airborne platforms, such as drones, balloons, and aerostats, have recently gained considerable interest in the communication sector. Free-space optical communication (FSOC) systems can deliver information wirelessly at high data rates (>1 Gb/s) using compact and lightweight optical terminals. Thus, airborne FSOC systems have been a recent subject of interest and activity, including emerging projects such as Google's Loon and Facebook's Aquila. One of major technical challenges of FSOC system is the stringent requirement of pointing, acquisition, and tracking (PAT) capability to ensure the reliable transmission between the transmitter and receiver, due to narrow beam width and movement of airborne platforms. By identifying and validating improvement techniques dealing with this issue, we developed simple models of propeller- and wind-induced vibrations and investigated the effect of those vibrations on the performance of FSOC systems operating in the stratosphere.

*(The page is intentionally left blank)*

# I. FSO communication

In 1880, 130 years ago, Alexander Graham Bell conducted the world's first wireless telephone between two buildings 200 meters apart [1]. The demand for wireless communication was growing especially during the world war in the early 20th century. The invention of lasers in 1960 was a promise to Free Space-Optics (FSO); however, it lost its momentum by the introduction of commercialized fiber optics installation. Free space optics may seem new fabled idea, but in fact, our day-to-day home appliances such as TV remote controls use Infrared spectrum for signal transmission.

Another major advance in FSO was made in 2001. MOSTCOM, one of Russian leader company in the field of FSO technology, started designing products ***Atrolink***. Artolink is a registered trademark of advanced products which are made via FSO technology." The product offers speeds 1Gbps, 10 Gpbs and 30 Gbps. This product is serially produced at aerospace enterprise, ***Ryazan State Instrument-Making Enterprise***. [3]

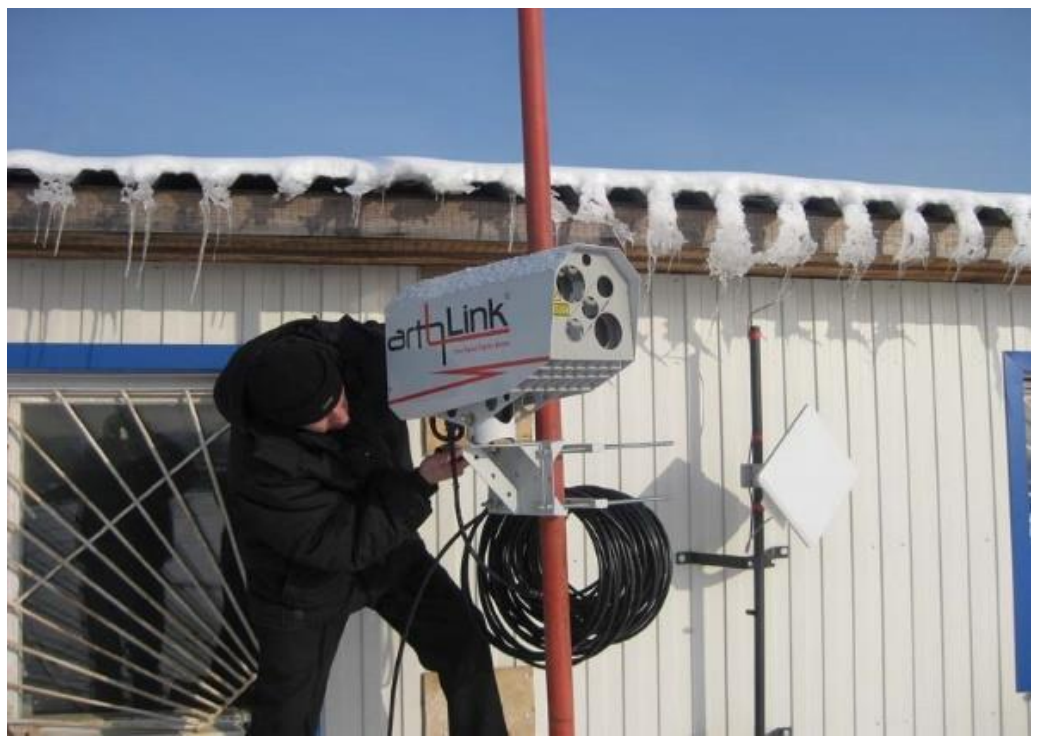

Figure 1. Atrolink device being maintained by a technician. [3]

The performance of Atrolink is remarkable, it have a capacity of reaching up-to a link speed of 30 Gbps in the full-duplex mode up to distance of 1.5 km. The device also maintains a low power consumption as low as 49W. For easy maintenance and management of the onsite device, it integrates UDP network in to its system. In addition, to keep the link connection in line of sight the device utilizes Autotracking. [3]

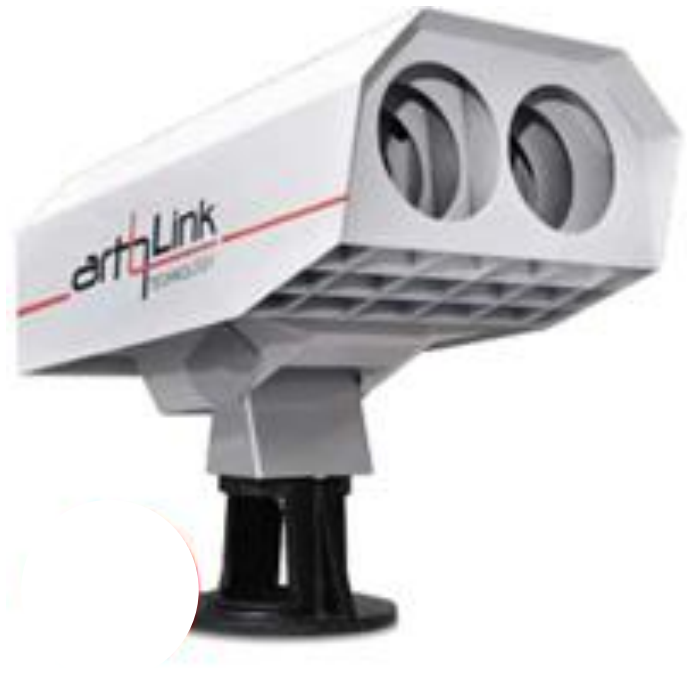

Figure 2. 30 Gbps ARTOLINK model, 'M1-30GE model'. [3]

The figure shown above spectacles the typical transceiver used in FSO systems in terrestrial FSO links. The digital signal-processing unit (DSPU) in the transceiver of an FSO system is more or less similar to RF system. However, there are considerable differences between FSO and RF in the method of transmitting and receiving the signal of interest. Since FSO's DSPU is similar with RF's DSPU, which are well studied and researched in RF systems in the past century, when studying FSO systems our main focus area of research is the method of transmitting and receiving an optical signal. The internal framework scheme of a typical FSO transceiver is shown in the figure below.

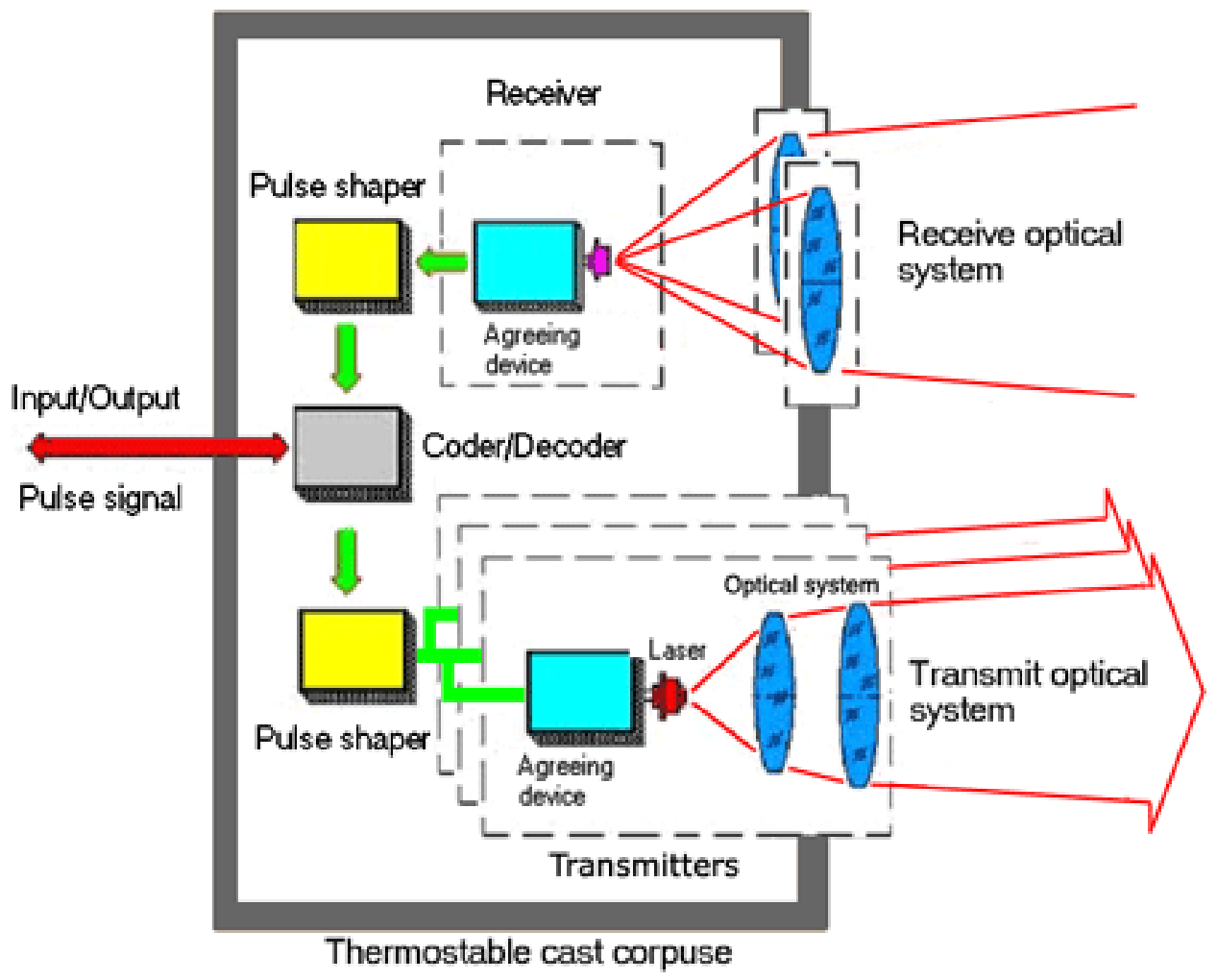

Figure 3. Transceiver model for FSOC of ATRO-LINK. [3]

In 2016, another company EC system improved ATRO-LINK to build a more reliable link, which can stand better in weather conditions. We believe this is the state of the art we are right now. [5]

Until this point, what we have reviewed so far is ‘FSO in terrestrial link’. The success of FSO in terrestrial links led to a parallel emerging branch of FSO encompassing High Altitude Platforms (HAPs). It involves spacecraft or balloons in the stratosphere as nodes for Free Space Optical Communication (FSOC) so-called ‘FSO in stratosphere link’. We believe the first optical communication test involving spacecraft was tested in 2008, just 10 years ago. Researchers at the German Aerospace Center succeeded in establishing communications between optic ground station and the TerraSAR-X Earth Observation Satellite [2].

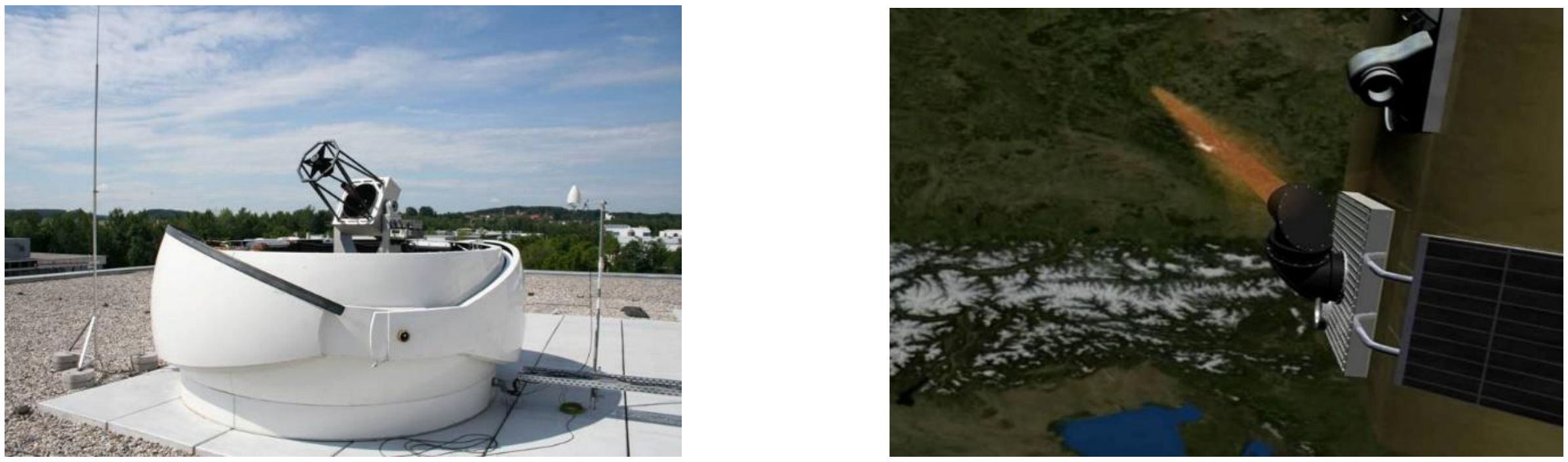

Figure 4. Optical ground station in Oberpfaffenhofen, Germany (left) and TerraSAR-X transmits data via laser beam (right). [2]

The station was able to track the satellite and receive signals from laser terminal on the TerraSAR-X satellite. The use of laser beams would enable data transfer from at speeds many times higher than those achieved at present, which would mean a vast step forward for wireless communications. [2]

In summary, FSO was introduced 50 years ago, whereas FSOC in stratosphere was introduced just 10 years ago. However, in these past decades, the actuality is the topic of FSO and FSOC have not been well assimilated and integrated in the realm of wireless communications, including the center of our research university, KAIST. We hope our research lead the light regarding the topic of FSOC and its importance in the building the future connected world. We anticipate this research to inspire students, researchers, scholars and scientists in electrical engineering and related fields to be exposed to the convenience and asset of FSOC in building a vast cutting-edge connected world.

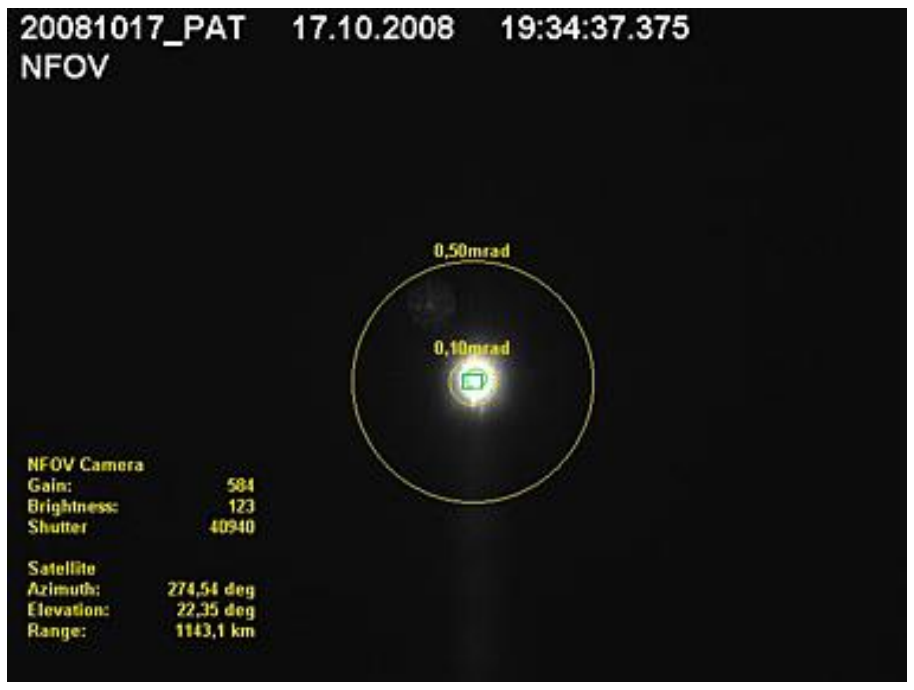

Figure 5. The laser signal transmitted back from space. [2]

# II. FSO communication in stratosphere

In the previous section of this report, we introduced FSO as a field of science and research where as in this section we will dig deep in a specific case of FSO, so-called FSO in stratosphere. It may be startling that more than a total of USD $400 million have been invested from 1998–2006 in the private sector primarily among four start-up terrestrial FSO based companies. Unfortunately, all four failed to deliver products that would meet telecommunications quality and distance standards. [4] The preeminent reason terrestrial FSO communication have been confined to non-commercial telecommunication is owed to the signal attenuation caused by the weather in the lower atmosphere. A FSO related military application research conducted by DARPA (Defense Advanced Program Research Agency), declassified report in 2011, supports our assertion that weather, particularly fog diminishes the performance of FSO by substantial factor. [6]

We have tried to touch upon terrestrial FSO and its limitations because of the presence of bad weather conditions for optical link in the lower atmosphere. One mechanism to constrain the attenuation incepted by the weather is to shift the medium of FSO to an elevated atmosphere layer above the dew point where weather conditions virtually disappear. This optimum layer is situated between the upper troposphere and the lower stratosphere. One significance of studying FSO in the stratosphere is, it will be able to compete with the current state of art wireless communication available in the present-day market, which is RF communication. Subsequently we will compare between the state of art of RF system and FSOC under Table [1]. The comparison table give us a clear insight on why FSOC is capable of outperforming RF communication; which is true if we ensure we accomplish adequate research and development employing a viable FSO system.

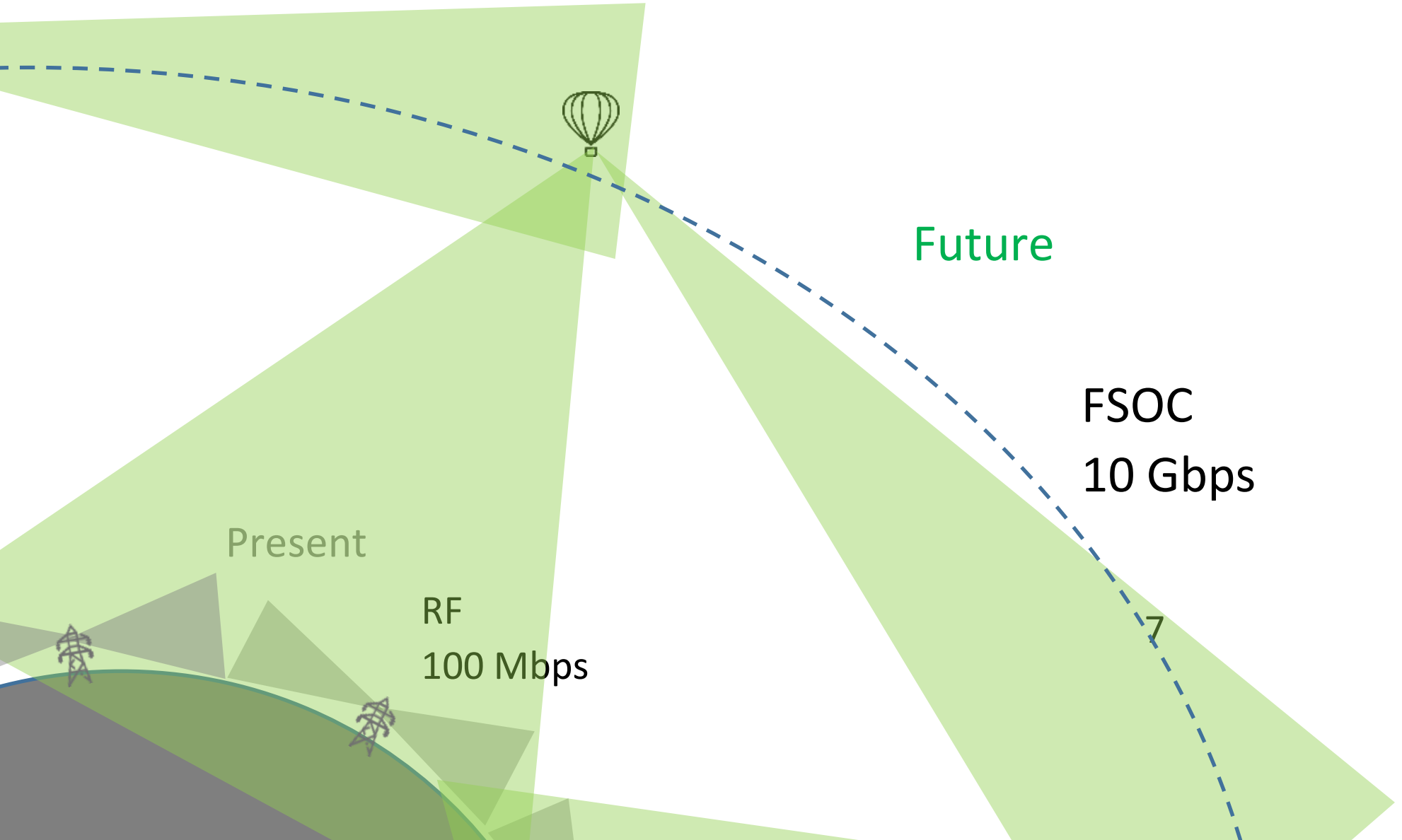

Figure 6. RF/MW communication Vs FSOC.

| | RF | FSOC |
|---|---|---|
| Emergency Period | Year of RF emerged ~ | 2000 ~ |
| Channel | Air, Liquid, Solid | Free-space, Liquid |
| Typical speed | Up to 100 Mbps | 10 Gpbs (with no fog) |
| Range | Varies from mm to 10 km | 100 km |
| Wavelength/ frequency | ~ 1cm/ 30 GHz | 1.5 $\mu$m/200 THz |
| Bandwidth limitations/ typical bandwidth/max bandwidth | Government Regulated << 30 GHz | Unregulated << 200 THz |
| Cloud Penetration | Good | Poor |
| Loss | 100 dB/km | 0.2 dB/km |
| Diffraction Angle | $10^{-2}$/D | $10^{-6}$/D |
| Transceiver gain amplifier | 10 dB | 50 dB |

Table [1]. Comparison table between RF and FSOC. This table helps us compare and contrast the two systems and to evaluate the limitation and advantages of FSOC compared to the existing widely used RF communication. Modified table from [6]

There are some important observation from the above table that are essential to discuss. Compared to RF, FSOC have low channel loss and wide available bandwidth with little or no regulatory limitations. As we can observe from table [1] the technical advantage of FSOC are considerable. This being said, it has the ability for scalable high-rate connections. [6]

A good example of FSOC that have been practically implemented are Project Loon and Project Aquila. Project Loon, which belongs to Alphabet's moonshot company X, floats balloons up to 20 kilometers (12 miles) above ground to send Internet service back down to rural areas. This service should, in theory, reach more people for less money than it would take to install base stations or fiber optic cables in those areas. [8]

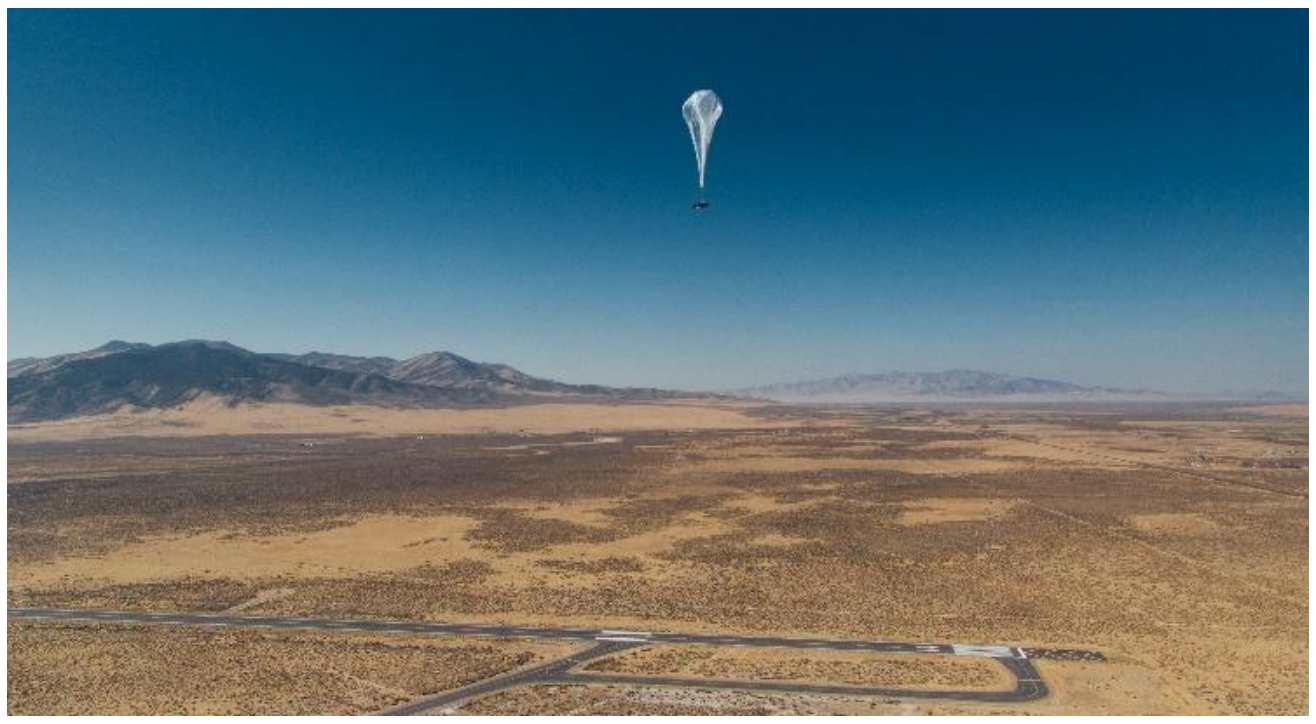

Figure 7. A Loon balloon on its way to Puerto Rico from Nevada. [8]

Salvatore Candido, the principal engineer for Project Loon as shown in figure 7, who led the development of the navigation system that controls the balloons, spoke with IEEE Spectrum about building it. He admits, "The one really hard thing is making the balloon be in the place you want it to be," Loon's launch engineers also had to figure out how to send massive balloons into the stratosphere. Each balloon has a solar-powered pump that can move it up and down by adding or releasing air. Adding air makes the balloon sink, because there is more mass spread out over the same volume, so it is heavier. With the help of that pump, the balloons can move into and out of winds that blow more or less constantly through the stratosphere. [8]

To create a navigation system that would coordinate the balloons' movements, the Loon team needed to understand stratospheric winds. They started with models and forecasts issued by scientific agencies including the U.S. National Oceanic and Atmospheric Administration. Then, they launched balloons to acquire their own data in areas they were likely to fly in. Candido articulates, "When we actually send balloons there that helps us build a more local model. One thing we really try and do is go and gather ground truth wind data in the places we want to serve." [8]

The other emerging FSO project is Project Aquila. Facebook announced a partnership with Airbus to expand ongoing development of its Aquila drone program, which the Silicon Valley-based company calls its "high altitude platform station (HAPS) broadband connectivity system." The two companies will be collaborating to "advance spectrum and aviation policy" and to demonstrate the "viability" of the HAPS concept. Earlier this year in 2018, Facebook's global aviation policy leader, David Hansell, delivered an in-depth overview of progress with the Aquila project at the annual Global Connected Aircraft Summit. Hansell said Facebook wants to use the solar powered drones to connect up to 66% of the planet's remote areas that feature outdated or no internet service at all. [9]

Unfortunately in midst of our research, a year after its launch. Facebook has announced that it will no longer pursue its dream of building a gigantic, solar-powered plane to blast internet to remote communities via laser. Astoundingly, it is just not practical. After the termination, Facebook announced, "Going forward, we'll continue to work with partners like Airbus on HAPS connectivity generally, and on the other technologies needed to make this system work, like flight control computers and high-density batteries. On the policy front, we'll be working on a proposal for 2019 World Radio Conference to get more spectrum for HAPS, and we'll be actively participating in a number of aviation advisory boards and rule-making committees in the US and internationally". [10]

Other ambitious projects developing HAP for FSO include UAVs Zephyr (Airbus) and Ion-drive airplane (MIT). Britain will buy the world's first high-altitude drone to hover from the stratosphere for months at a time. The solar-powered Zephyr aircraft will be used by Special Forces and regular soldiers as part of a £2 billion boost to intelligence-gathering capabilities, from report of defense. The Ministry of Defense is to spend £10.6 million on two prototypes to be built in the UK. Test flights are expected next year in 2019. Michael Fallon, the UK defense secretary, said. "They will be able to fly higher and for longer to gather constant, reliable information over vast areas," The aircraft (Zephyr) known also by it pseudonym high altitude pseudo-satellite flies at 21 km, nearly twice the height of a commercial airliner, weighs only 30kg and travels at about 48km/h. [11]

The subsequent HAP we will discuss here is Ion-drive airplane shown in figure 8. Nature journal reports researchers at the Massachusetts Institute of Technology (MIT) in Cambridge led an aviation breakthrough that will draw inevitable comparisons to that shaky and fragile first journey by air. The aeroplane is powered by a battery connected to a type of engine called an ion drive that has no moving parts. [12]

There are no passengers, either. The whole device, which has a 5-metre wingspan, weighs just 2.5 kilograms, about one-tenth of a typical commercial flight passenger's baggage allowance. Currently, the aeroplane hardly gets off the ground, cruising in tests at an altitude of 0.47 metres. However, with amplified R&D, anyone who watches the machine fly can surely see glimpses of a future with cleaner and quieter aircraft. [12]

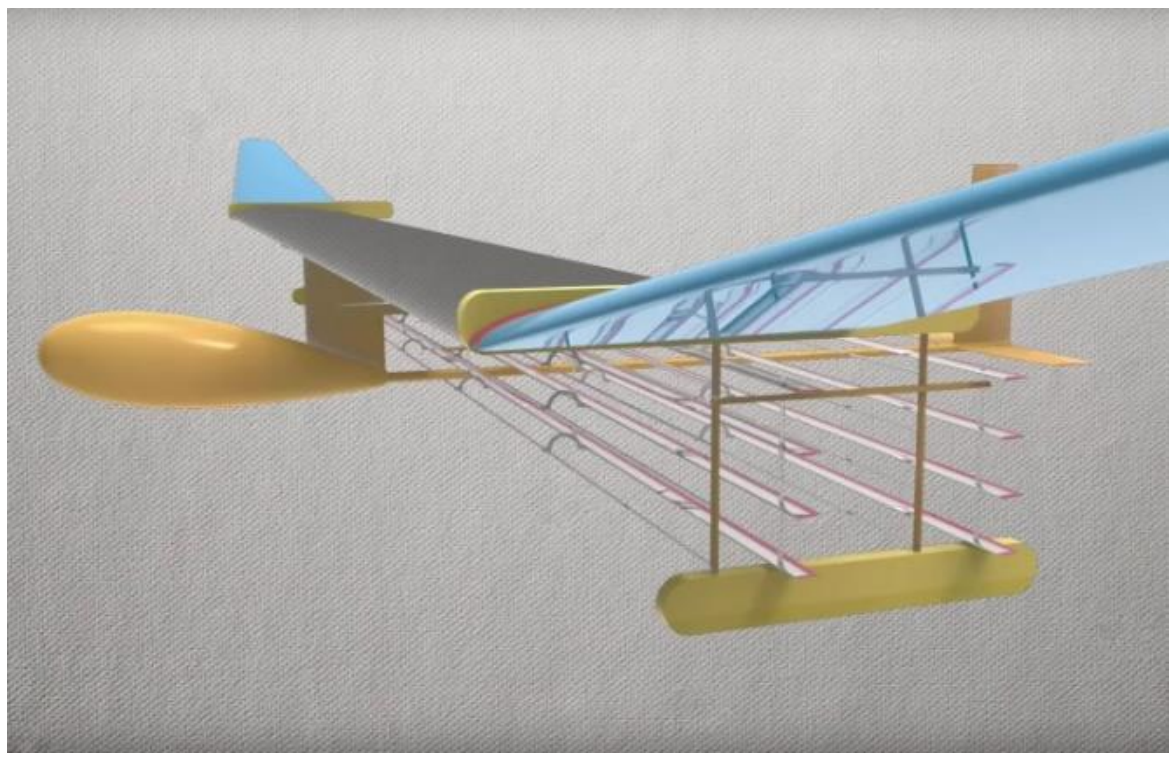

Figure 8. Prototype of Ion-drive glider. [2]

The news article (Nature Journal) examines into the technical details and the challenges that must be addressed to scale up this prototype plane. Is such a goal achievable? The aircraft with ion-drive create thrust by using electrical forces to accelerate ions in a fluid to form an ionic wind. The thrust is produced only by the wind generated by the movement of ionized air molecules as current passes between two electrodes, one thinner than the other. [12]

Ionic wind was first identified in the 1960s, but most scientists and aviation professionals since have insisted that the process was never going to be efficient enough to be useful, However, MIT researchers demonstrate the first flight of an aeroplane propelled in this way, they also show that the efficiency will increase as the velocity of the aircraft increases, because the electrodes that act as the engine create such little aerodynamic drag. [12]

The scientists' success will surely spur on others to re-explore a technology that was long forgotten. This will no doubt include military research, and some of the possible applications silent drones and engines with no infrared signal that are thus impossible to detect. This includes our research, if un-shaky Ion-drive proved to be practical; the pointing error of FSO HAPs due to propeller-induced vibration could be easily circumvented. Ion-drive engines are also much-needed option to improve the efficiency and environmental impact of aircraft engines.

FSO in stratosphere could have multiple implications, in this report we will see the leading areas examples where FSO stratosphere systems could be deployed.

- *FSO for remote areas.*
- *FSO city-to-city link.*
- *FSO future cities integrated link.*
- *FSO interplanetary communication link.*

FSO deployment in remote areas could speed the pace of the economy of developing countries, such as Africa as shown in figure 9. FSO connected zones implies, full coverage of internet and abundance of information that have the capacity to revolutionize third world countries by improving education and enabling business connection with the outside market.

FSO could also be deployed to connect cities in developed countries. Developed countries will be able to provide internet speed up to 100 times than the current capacity of wireless communication. Figure 10, below shows the potential scope of expanding FSOC between industrialized cities in the near future.

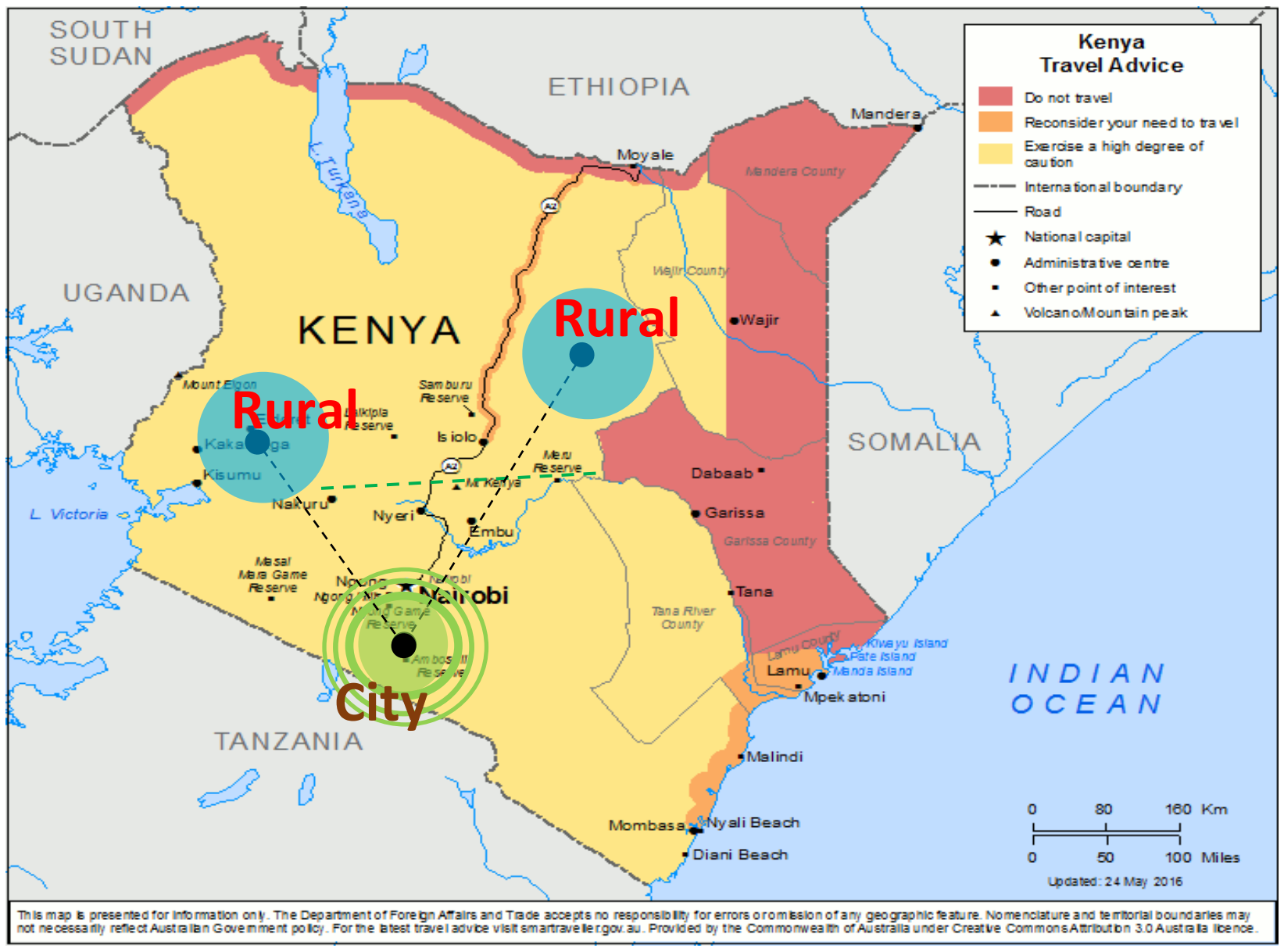

Figure 9. An example of FSO deployment in Africa.

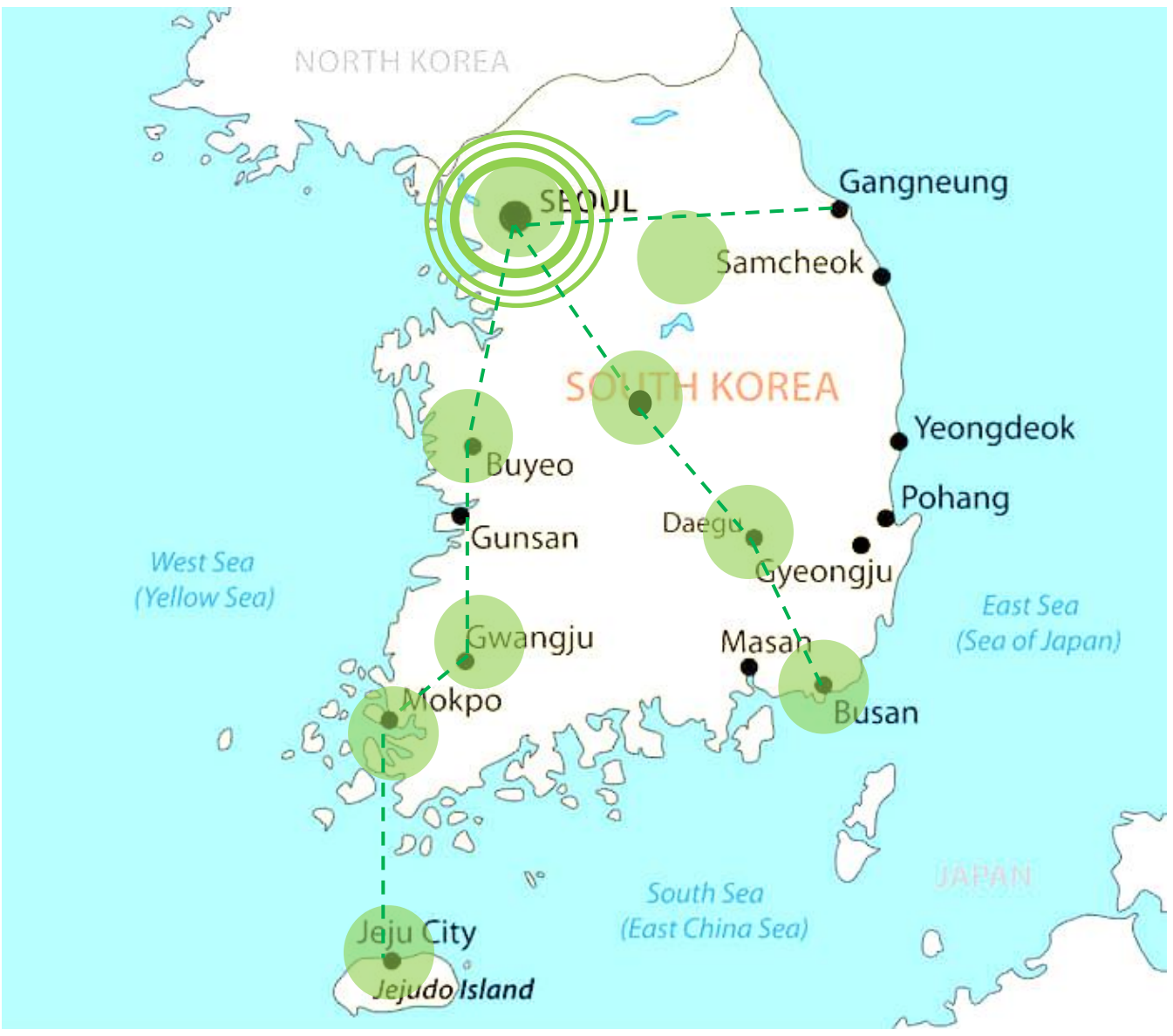

Figure 10. An example of FSO fast internet 30 Gpbs 2030 VISION for developed countries, (with maximum of 100 ~ 150 km link range between cities).

FSO also has the potential to connect future technology cutting-edge metropolitan cities such as Seoul- New York-Tokyo 'Hyper Speed Secured FSO link' as illustrated in figure 11. In addition to future cities, HAPs can also be used for deploying interplanetary communication link for fast speed communication, as shown in figure 12, in the coming decades among different planetary stations

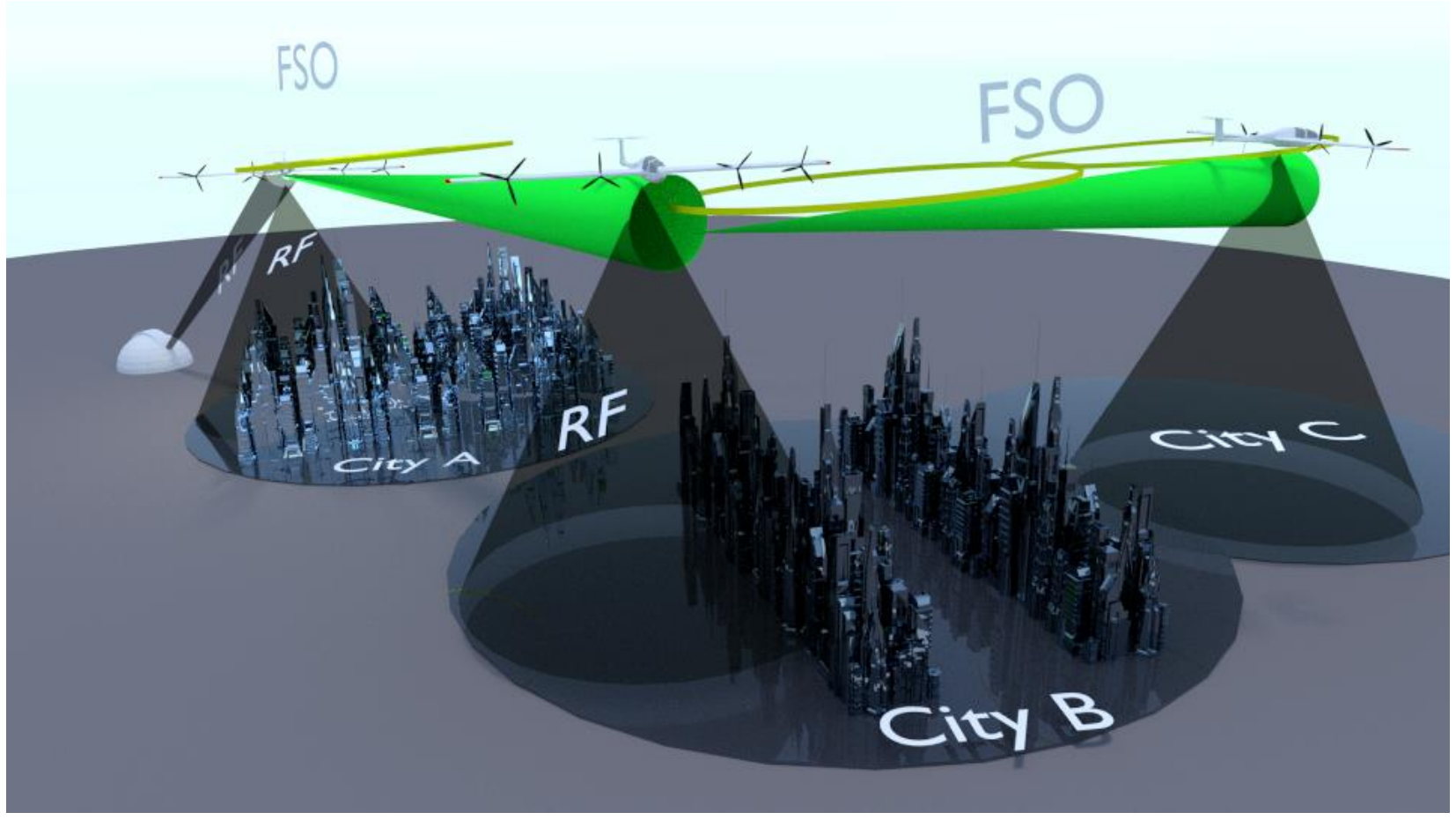

Figure 11. The future of FSO in the near future.

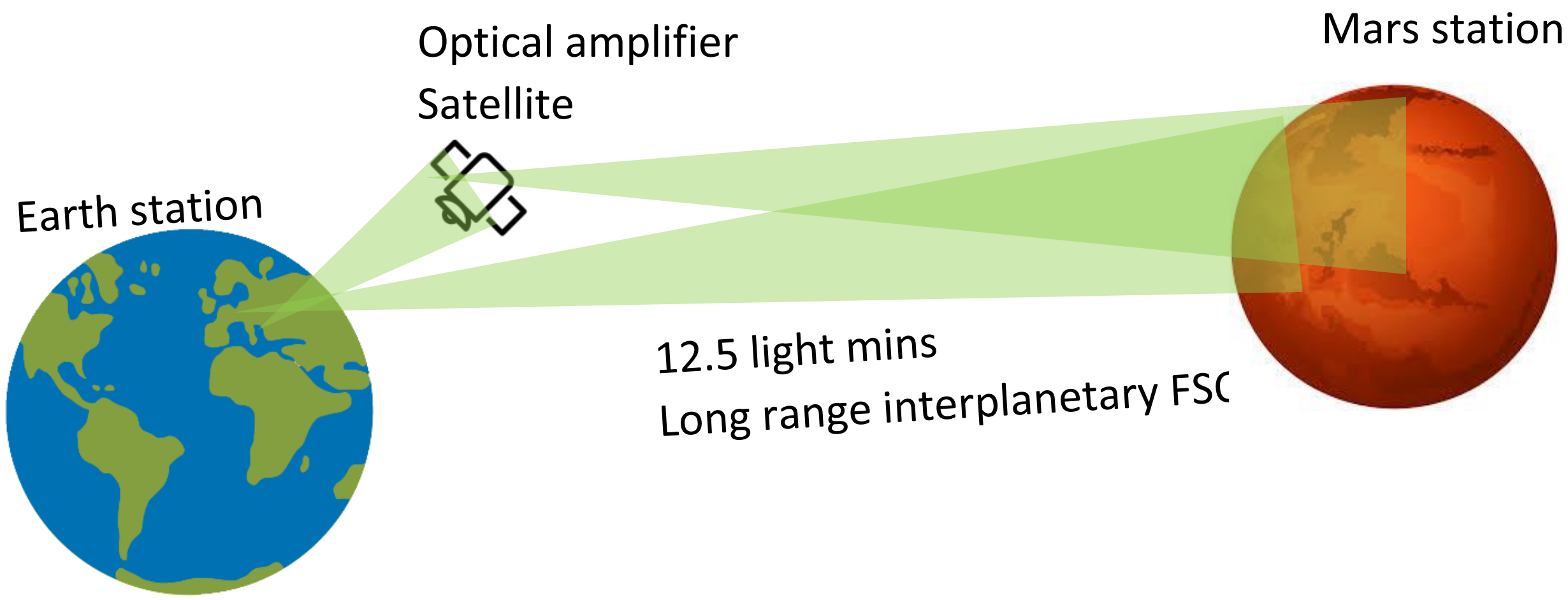

Figure 12. Interplanetary FSO communication link.

# III. Causes of attenuation in FSOC

Why do we study the power attenuation of FSO systems? Even though FSOC promises us incredible data speeds. The reliability of FSO units has been a problem for commercial telecommunications. Independent studies find too many dropped packets and signal errors over small ranges (400 to 500 meters) when fog is present. [4] These studies agree the stability and quality of the link is highly dependent on atmospheric factors such as rain, fog, dust and heat. One of the big reason the FSOC momentum has slowed is the past decades is factor of weather condition, which could bring down the speed from Gb/s to few Mb/s.

The transmission of FSO from transmitter to receiver is illustrated in figure 13. The attenuation of the FSO link follows between the transmitter and the receiver blocks. The main causes for attenuation are highlighted in figure 13, which include atmospheric loss, geometric (space) loss and pointing (misalignment) error. We will discuss these in the following section.

Attenuated power of FSO is defined as the ratio of received and transmitted power as in equation (3.1) or as the difference between the two powers in dB scale, as in equation (3.2). We will now see the each attenuation factors of FSO one by one.

$$Attenuated\ power\ = \frac{Received\ Power}{Transmitted\ Power}\ (Linear) \quad (3.1)$$

$$Attenuated\ power = \ Recieved\ power - Transmitted\ power\ (dB) \quad (3.2)$$

$$\text{Received Signal} = (P_T)(G_T)(\tau_T)(\tau_{ATM})(\lambda/4\pi L)^2(\tau_{Pointing})(\pi D/\lambda)^2(\tau_R)$$

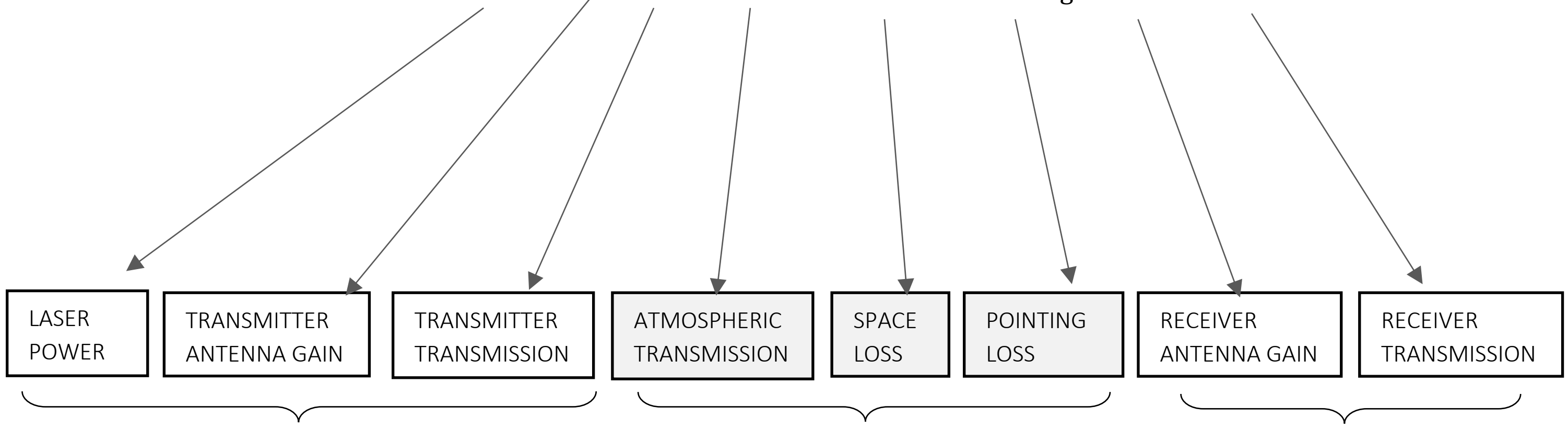

Figure 13. FSO Link from transmission to reception. Modified diagram from [7]

A. Weather conditions

Weather conditions affect FSO link undesirably when atmospheric factors such as rain, fog, dust and heat is present in the atmosphere channel. [4]. Weather attenuation of FSO link could bring down the speed drastically from remarkable few Gb/s to undesirable slow connection of few Mb/s. Fortunately, FSO link in stratosphere is not affected weather since the stratosphere layer is found at much high altitude relative to the dew point. Hence, owed to this reason we limit our scope and choose not to discuss the effect of weather conditions in FSO stratosphere link between HAPs.

B. Geometric loss

Geometric loss is attenuation of FSO link caused by the part of laser beam not striking the receiver aperture surface, as depicted in figure 14. Hence, geometric loss is defined as the ratio of received power by the aperture to transmitted power at the receiver, given by equation (3.3) and equation (3.4) in log scale. Since power is inversely square proportional to area of the beam, we can trivially obtain equation (3.5) and through substitution of area estimation referring to figure 13, we obtain can verify equation (3.6). [1]

[1] In verifying equation (3.6), we use small angle approximation of the beam divergence angle. This can be verified because the beam angle is very narrow and the link range in large magnitude.

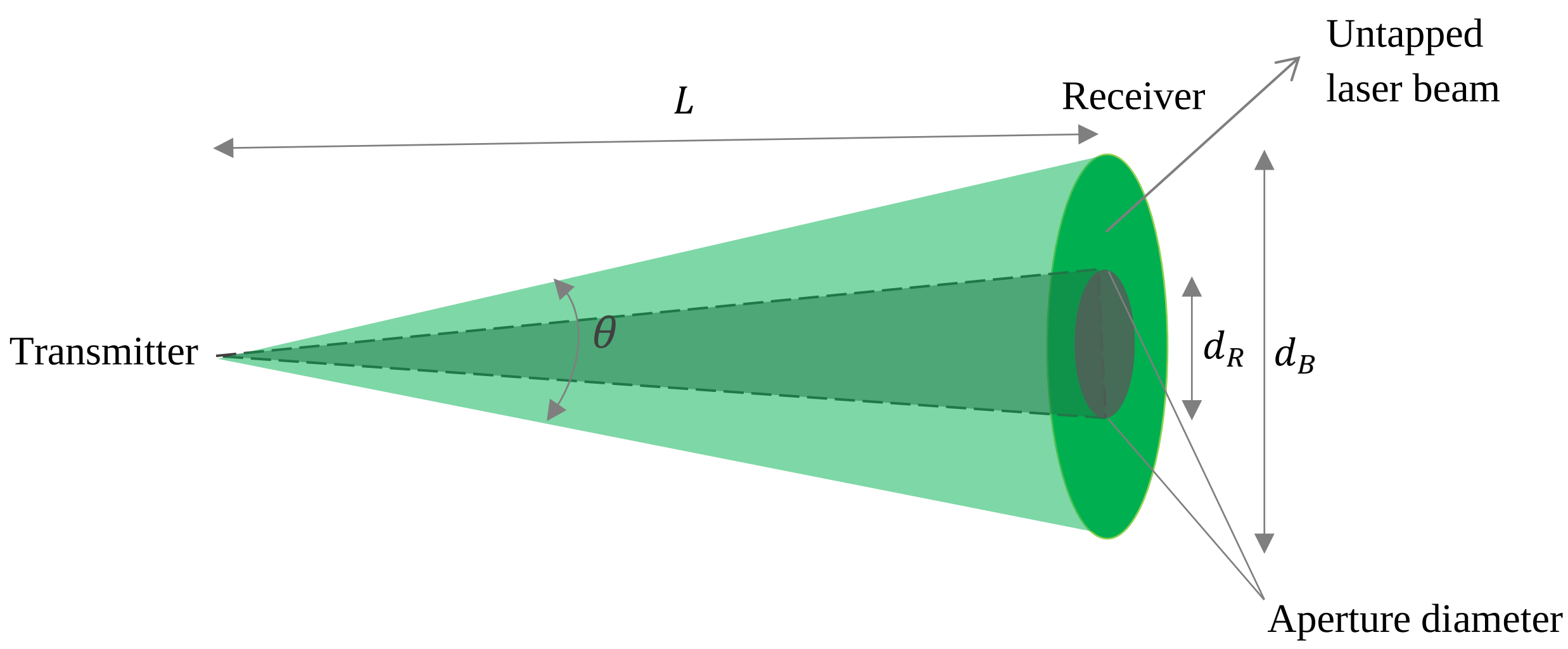

Figure 14. Illustrating geometric loss in typical FSO link

$$Geometric\ Loss\ [linear] = \frac{Received\ power}{Transmitted\ power} \quad (3.3)$$

$$Geometric\ Loss\ [logscale] = 10\log\left(\frac{Received\ power}{Transmitted\ power}\right) \quad (3.4)$$

$$Geometric\ Loss\ [logscale] = 10\log\left(\frac{Reciever\ lens\ area}{Transmitted\ beam\ area}\right) \quad (3.5)$$

$$Geometric\ Loss\ [logscale] = 10\log\left(\frac{\pi {d_R}^2}{\pi(\theta L)^2}\right) \quad (3.6)$$

In our research introduction, discussing about emerging FSO projects we covered about Project Loon. To help us more understand more about geometric loss we will estimate the geometric loss with Google Loon's FSO test flight parameters, as illustrated in figure 14.

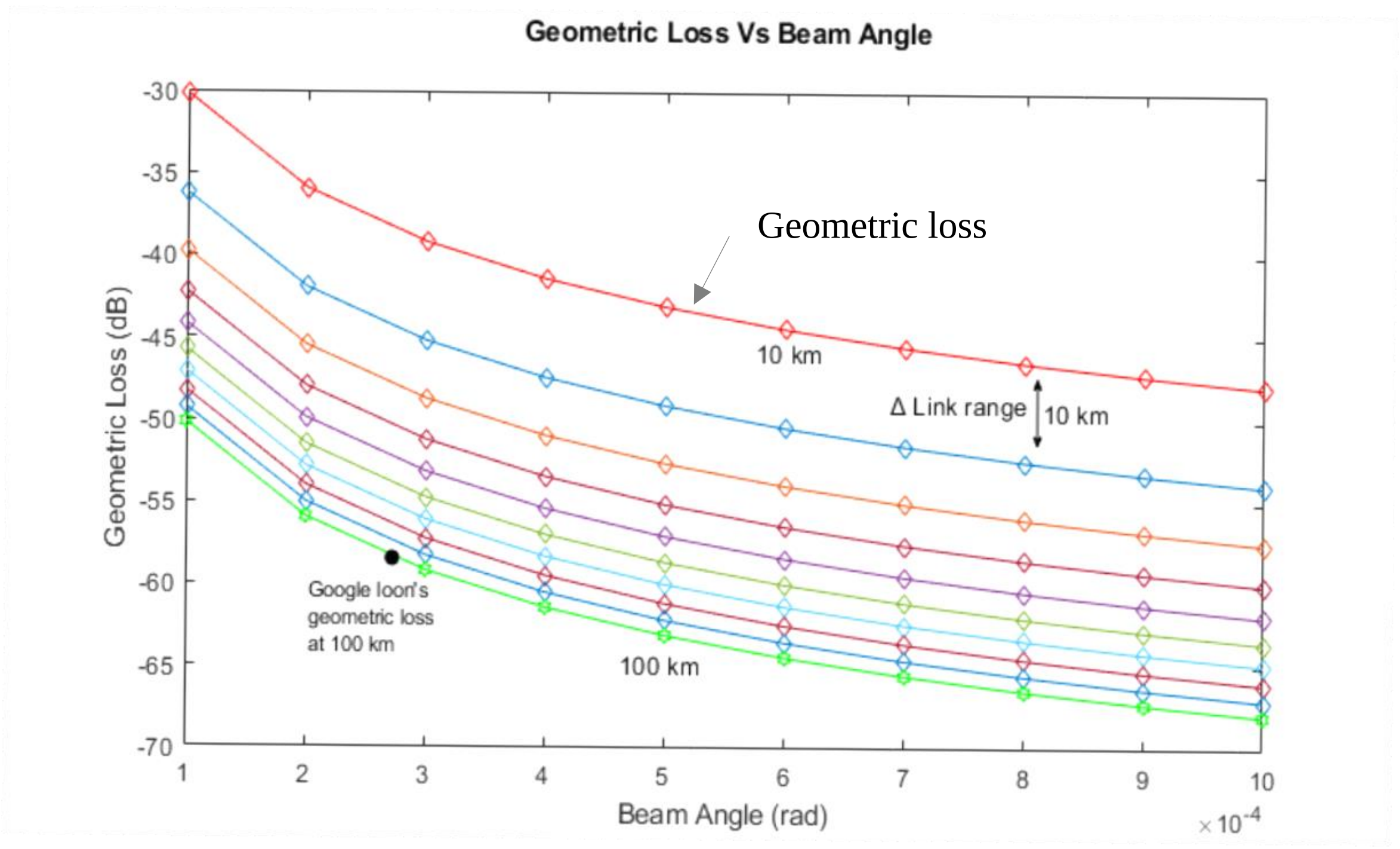

Figure 15. The figure shows the varying geometric loss for 0.037m transmitter and receiver aperture, with varying distance (link range) between the HAPS. The top curve corresponds to 100km link range and the bottom curve to 10km link range.

## C. Misalignment error

Misalignment error is the final attenuation factor we will cover in this report. Misalignment error, as its name gives clue is the power attenuation caused due to the misalignment of the transmitter and the receiver HAP in an FSO link, as shown in figure 16. Since misalignment is the focus of our research, we will study it in detail in the following section.

# IV. Misalignment error

Unlike radio communication systems, FSOC (Free Space Optical Communication) relies on line of sight (since light travels in straight-line). Henceforth, optical alignment of transmitter and receiver turns out to be extremely critical in FSO system design. FSO transceivers including those used in terrestrial link use real time auto-tracking to compensate movement that is caused by thermal expansion of mast and motion of the building caused by wind. [5] If FSO transceivers in terrestrial link prerequisite auto-tracking, it is logical to draw FSO transceivers in stratosphere, which are exposed to much stronger speeds and propeller vibration (in case of UAVs) need an enhanced auto-tracking system.

To provide better auto-tracking system for FSO system design. We need to understand the vibration parameters involved in the FSOC, which is what our research principally focuses about. In this report, we study the vibration of HAPs, which solely carry the FSO transceivers in the stratosphere. We presume the vibration of HAPs is dominantly affected by strong winds and propeller of HAPs. During the time of our research, there have not been significant research in the vibration of HAPs and alignment of FSOC in stratosphere. Hence, we hope our research becomes a guideline, for students, scientists, researchers or companies who are interested in studying FSO alignment in stratosphere.

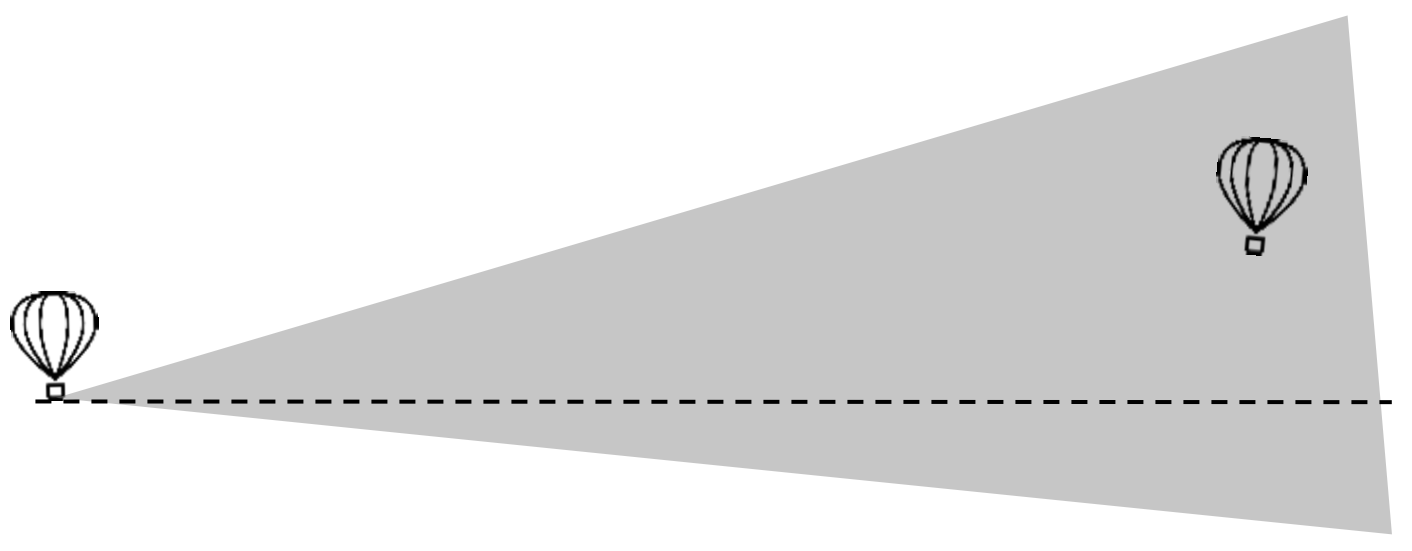

Figure 16. Misalignment error between airborne HAPs.

While doing survey on the subject of misalignment error, we identified that there is no available satisfactory research conducted regarding misalignment error of FSOC.

Henceforth, we were motivated to do our version of experiment by collecting pointing error data and use the data to understand further about pointing errors and their relation with vibration of FSO transceivers.

A. Experiment results

In this experiment, our objective is measuring the pointing error distribution in free space at the receiver of the FSO link, as shown in figure 17.

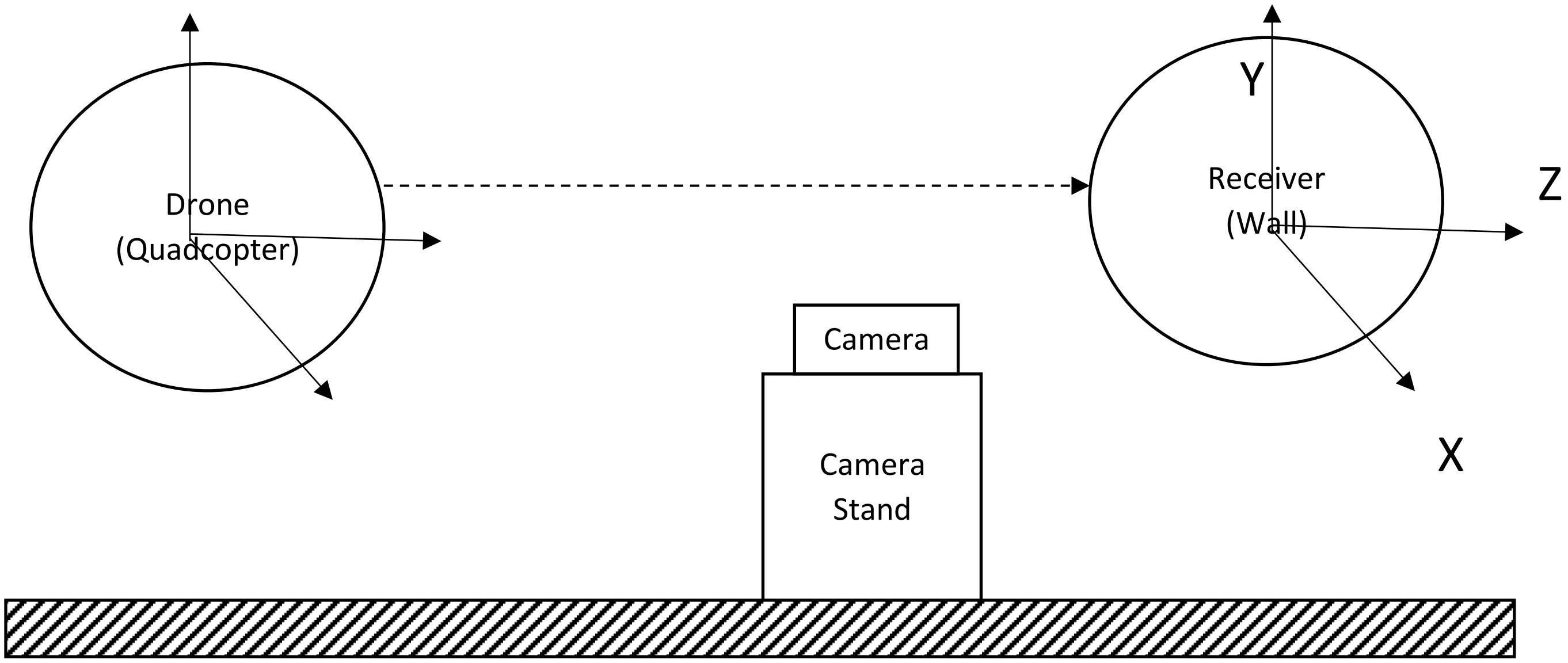

Figure 17. Experiment setup for measuring the pointing error induced by propeller vibration of a drone.

The pointing error statistical data we estimated after gathering real-time data at the receiver from the experiment is presented in figure 18. Where ‘X’ and ‘Y’ represent the linear vibration magnitude, ‘$\theta_x$’ and ‘$\theta_y$’ represent the estimated angular vibration of the drone, and ‘$\rho$’ represents the magnitude of ‘$\theta_x$’ and ‘$\theta_y'$. [2] Table [2] illustrates the figure in table form using probability parameters.

[2] In calculation of ‘$\theta_x$’ and ‘$\theta_y$’we presume the angular vibration of the transmitter is the dominant cause of lateral vibration at the receiver and the effect of linear vibration of the drone is relatively small.

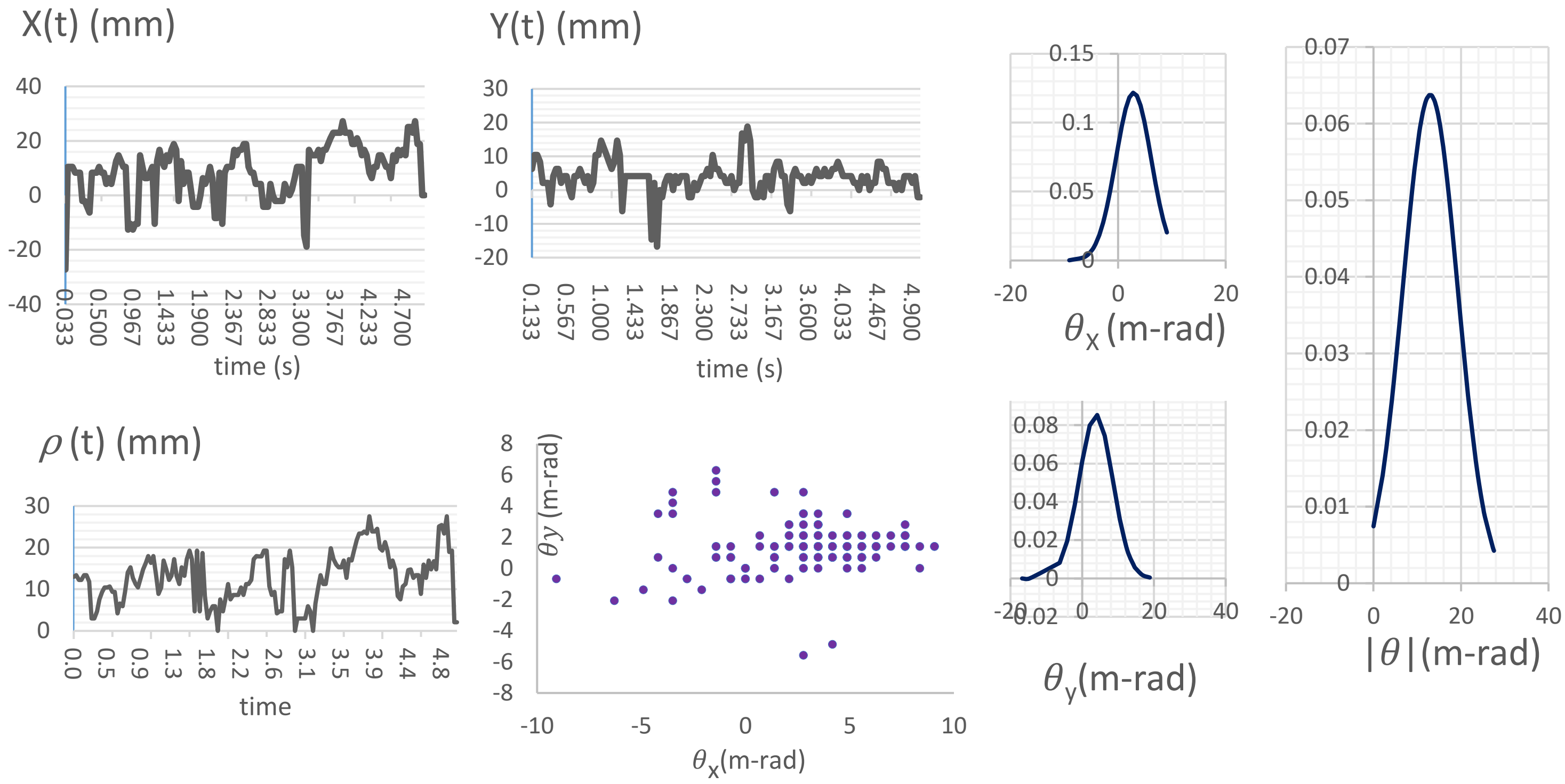

Figure 18. Pointing scatter plot over a 5 sec flight interval, statistical data gathered from our pointing error experiment.

| Statistical Parameter | $\mu_{\theta_x}$ | $\mu_{\theta_y}$ | $\mu_{\lvert\theta\rvert}$ | $\sigma_{\theta_x}$ | $\sigma_{\theta_y}$ | $\sigma_{\lvert\theta\rvert}$ |
|---|---|---|---|---|---|---|
| Value (mm) | 2.8749 | 1.2746 | 12.9700 | 3.2768 | 1.5535 | 6.2563 |

Table [2]. Statistical pointing error data table from the experiment.

We will now elaborate the result plotted in figure 18. ‘X (t)’ and ‘Y (t)’ Vs time graph represent the lateral pointing error vibration in ‘X’ and ‘Y’ dimension respectively. By observing the magnitude variation of those two graphs closely, we can propose ‘X (t)’ vibration is shakier compared to ‘Y (t)’. ‘$\rho(t)$’ Vs time represents the vector magnitude of the pointing error lateral vibration at the receiver. ‘$\theta_x$’ and ‘$\theta_y'$ are obtained from ‘X (t)’ and ‘Y (t)’ utilizing small angle approximation and presuming angular vibration have the dominant effect in misalignment error. The plot ‘$\theta_x$’ Vs ‘$\theta_y$’ represents the pointing error distribution induced by the angular vibration of the transmitter drone. The distribution plot is more scattered along the ‘X’ axis signifying ‘X’ vibration is shakier compared to ‘Y’. Based on this data we plot the PDF ‘$\theta_x$’, ‘$\theta_y$’and ‘$\rho(t)$’.

As we have emphasized, accurate pointing is critical to link closure. During the flight we tracked the instantaneous beam position, which we represent as a two-dimensional vector $\theta = (\theta_x, \theta_y)$, where $\theta_x$, $\theta_y$ are orthogonal components of the angular position. Suppose the anticipated beam position corresponds to (0, 0) so that $|\theta|$ is the angular pointing error.

Figure 18 (middle bottom plot) illustrates a scatter plot of $\theta$ and corresponding PDF (probability distribution function) of $\theta_x, \theta_y$ and $|\theta|$ for a representative 5 second interval of the flight. Referring from table [2], orthogonal components of the pointing error had mean and standard deviation on the order of 2.8 mrad and 3.3 mrad, respectively. The angular pointing error had mean and standard deviation on the order of 12.9 mrad and 6.3 mrad, respectively. We illustrate Gaussian fits to the orthogonal components and a Rayleigh fit to the angular pointing error. In the next section, we will discuss how we obtained the results shown in this section.

## B. Experiment procedure

In this section, we will construe in depth how we obtained the experiment result presented in the previous section and systematically clarify each experiment procedure. Firstly, to establish the objective of the experiment, the motivation for this experiment sparked when we sought to conduct a low-cost UAV FSO link system and study the pointing error in the FSO link. Our first phase was to design a low-cost FSO system analogous to the practical FSO problem, shown in figure 19. Hence, considering a realization of low-cost system and intention of fundamental study of the subject, we settled for drones hovering at the lower surface earth's atmosphere instead of typical UAVs systems hovering in stratosphere.

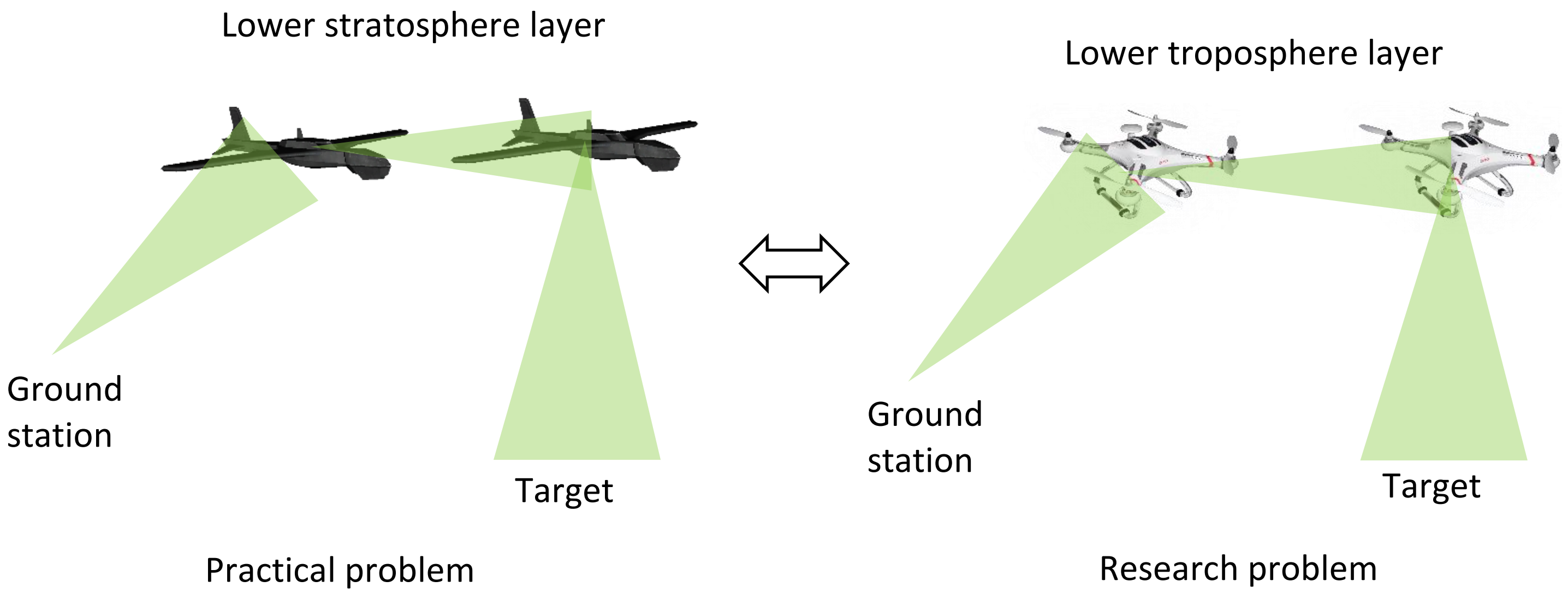

Figure 19. Designing low-cost FSO system akin to the practical problem at hand.

Following up from figure 19, to line up with our research objective, we isolate the FSO link between the two HAPs as shown in figure 20. To remind our research objective, our aim of the experiment is to study FSO link misalignment error induced by vibration. Going back to figure 20, we have two Elevations; Elevation B will used as a control variable with no vibrations and at Elevation A we collected the data with the presence of vibration.

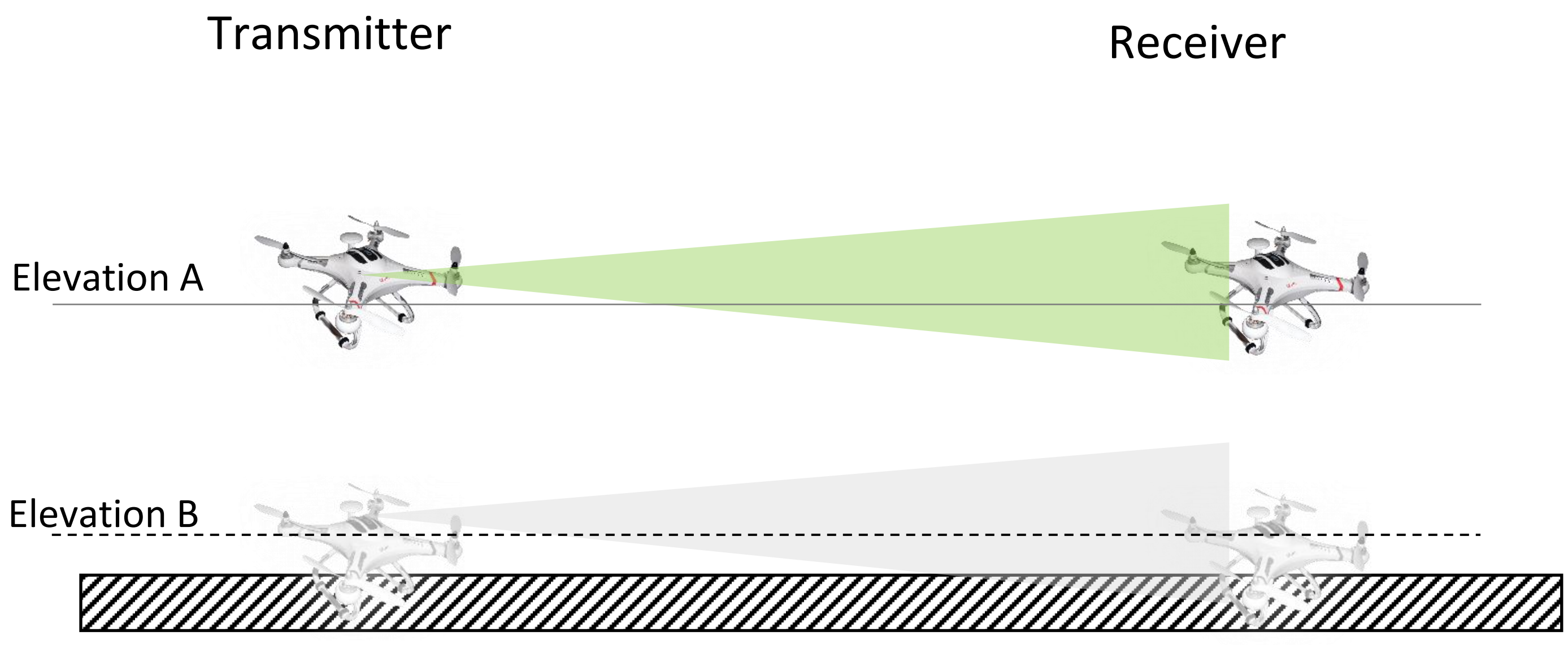

Figure 20. Motivation for the experiment.

Finally, to do a fundamental experiment we will start by studying 'one-way FSO link' (meaning one transmitter and one receiver). This choice eliminates the need for requiring another drone, since in one-way FSO link the receiver we want to study is presumed stationary, we can choose arbitrary object such as wall, as depicted in figure 21. Hence, for our experiment we treat the drone as the transmitter and the wall as a receiver. The transmitter for our experiment is a quadcopter drone 'Holy Stone GPS FPV RC Drone HS100 '. The drone is knotted with a laser pen to be used as a laser source, shown in figure 22. The receiver in our case is a block wall that casts a laser beam footprint in dark, shown in figure 23. Consolidating the depicted transmitter and the receiver, the experiment setup schemata is well-illustrated figure 24, whereas a picture of the experiment setup, which took place inside a hall, is shown in figure 25.

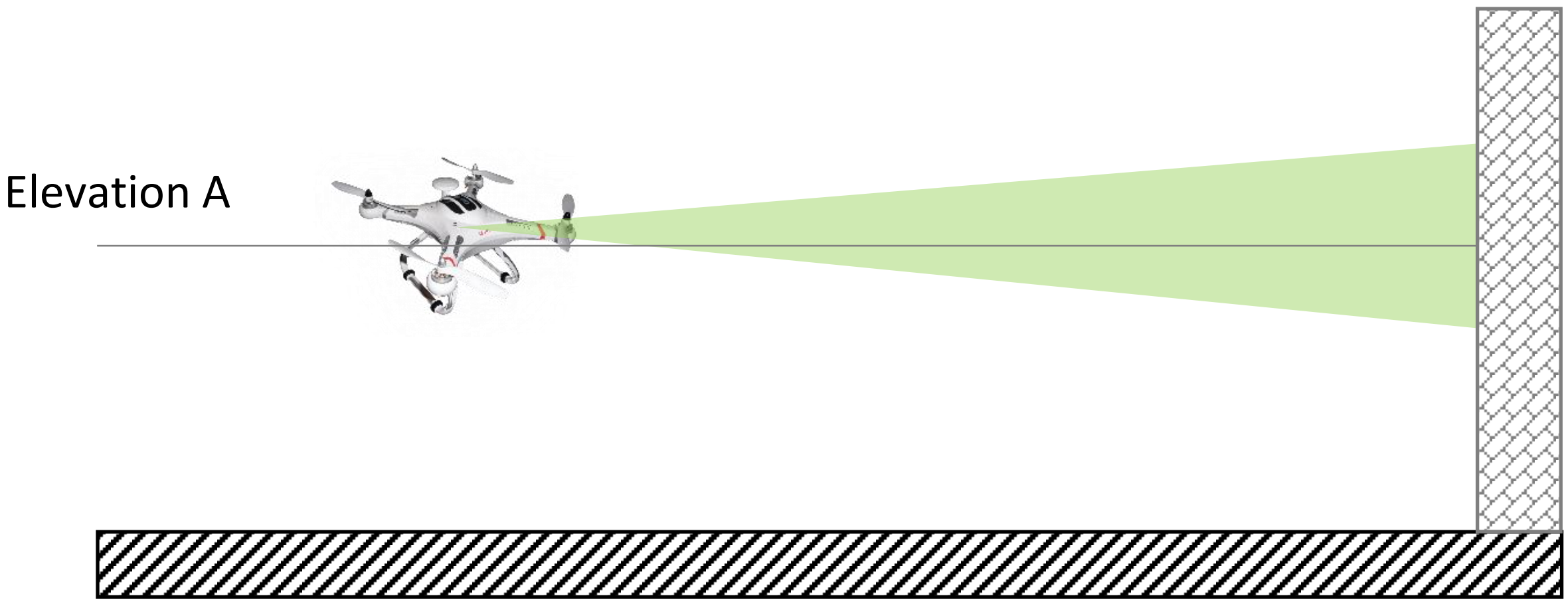

Figure 21. FSO link experiment method.

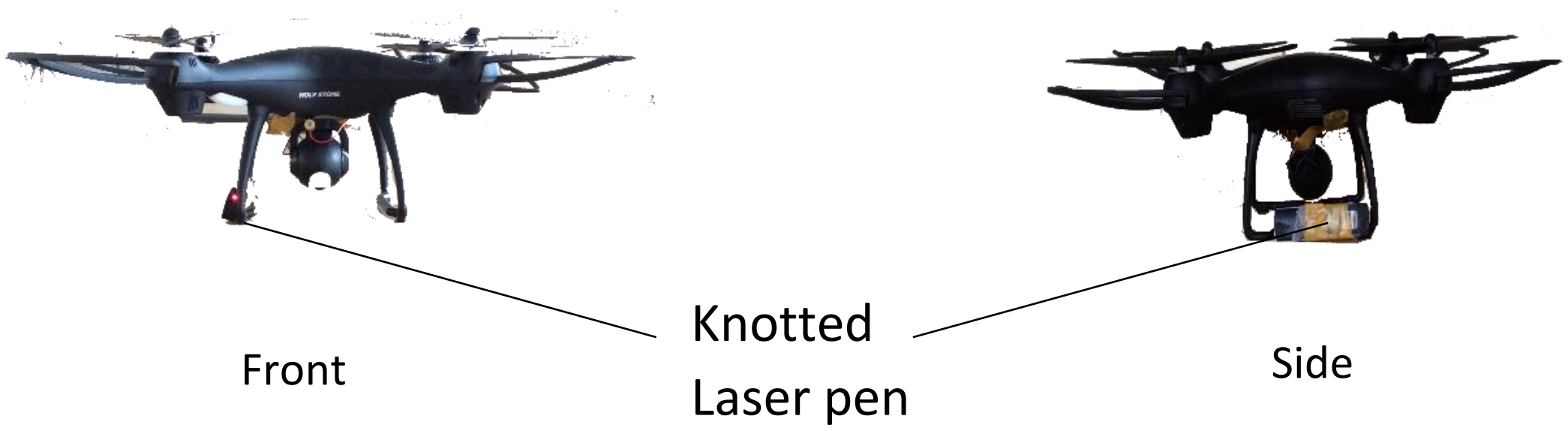

Figure 22. Transmitter of FSO link knotted with 610 nm laser pen.

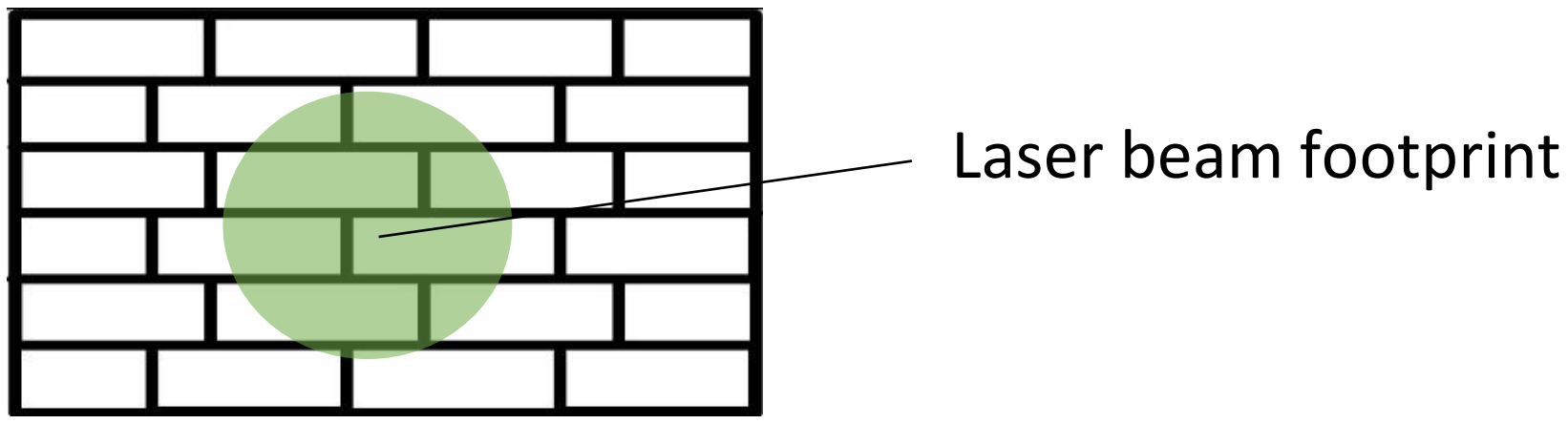

Figure 23. Receiver of FSO link.

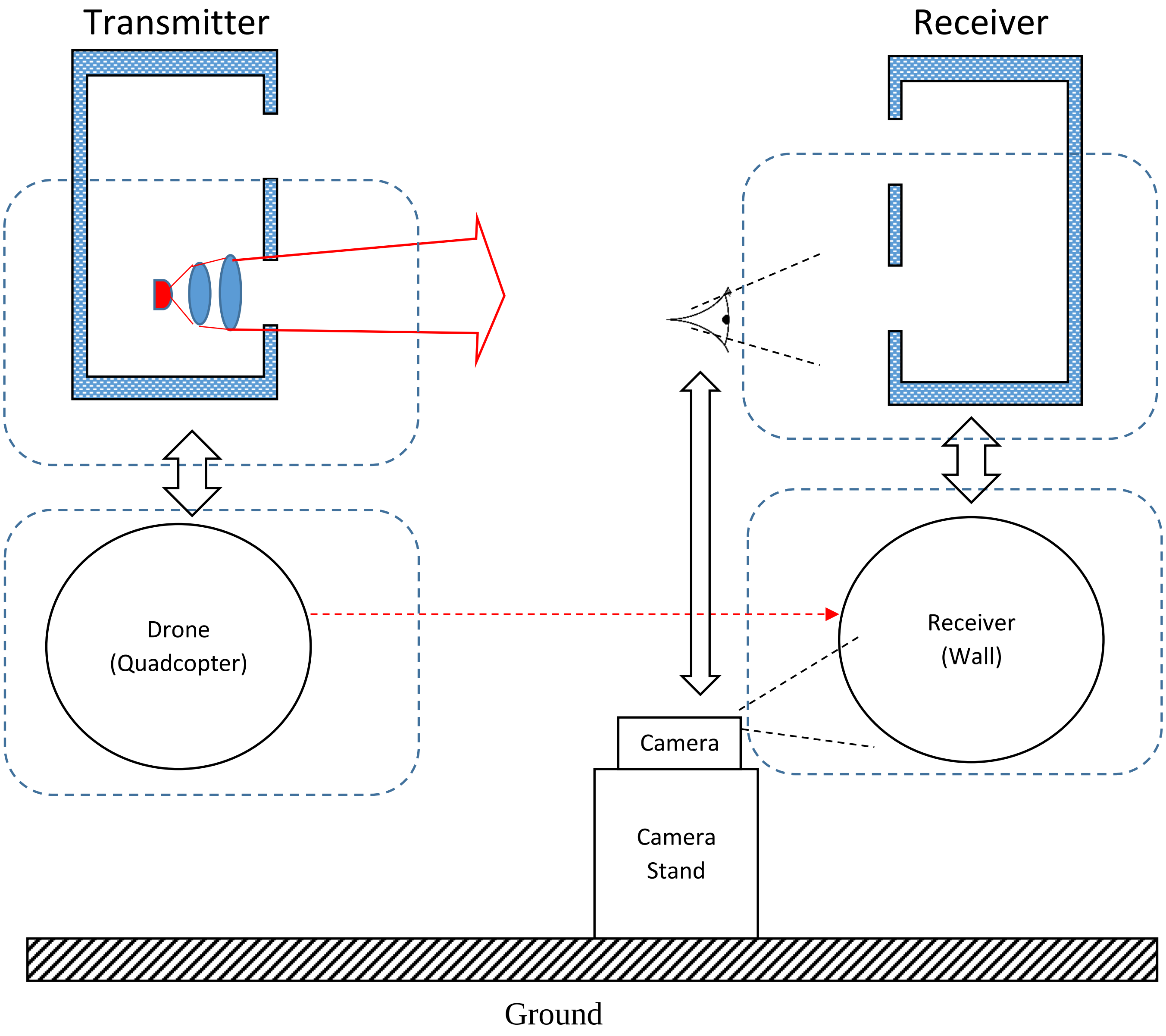

Figure 24. FSO corresponding of our misalignment error experiment setup.

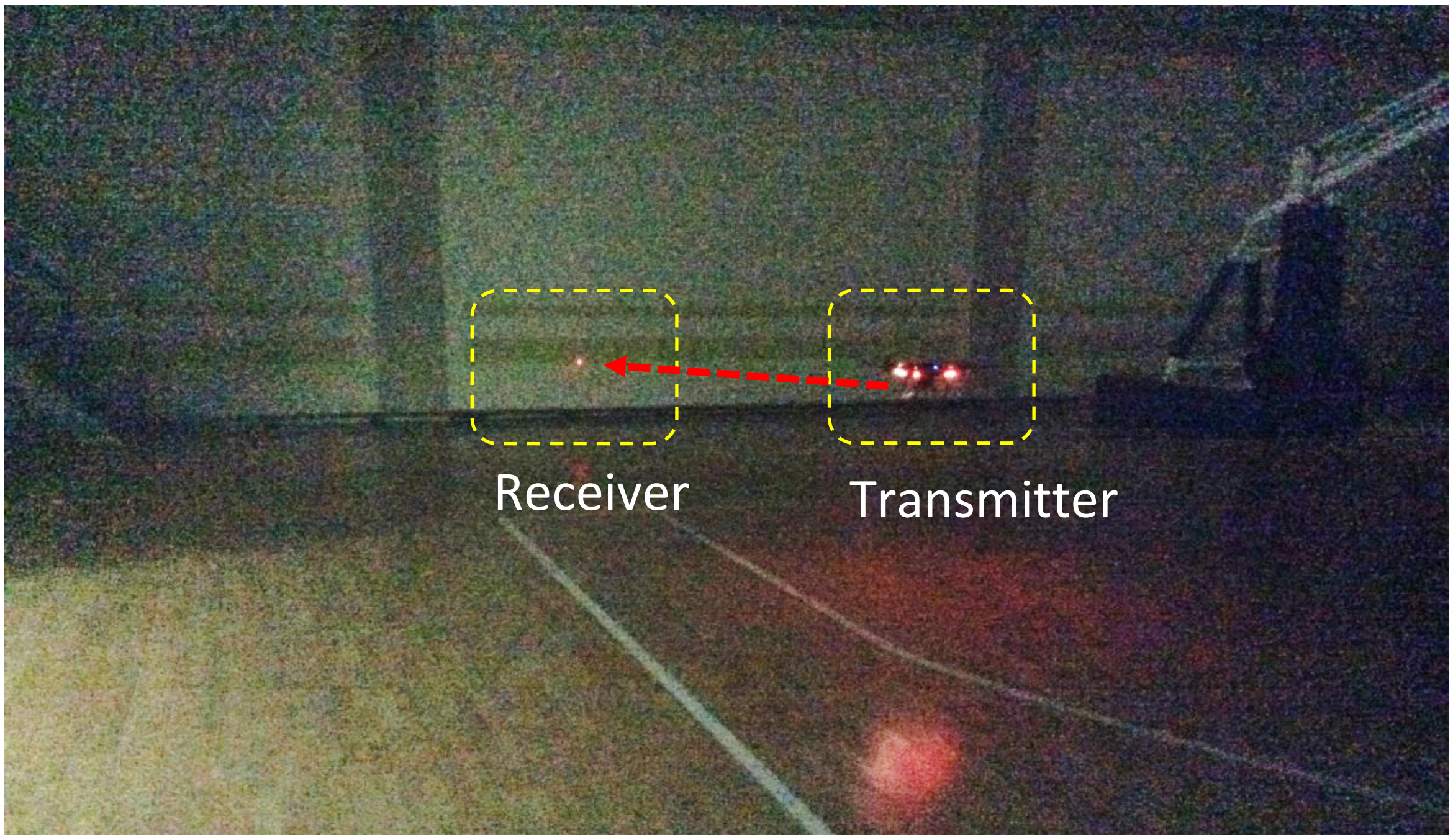

Figure 25. Experiment setup.

Measuring the pointing error entails pointing and tracking (PAT) the laser beam footprint at the wall. To measure the pointing error swept across time on the receiver wall, we utilize a video recording device ('in our case iPad air camera recording at 30 fps). After recoding the video, we measure the pointing error pixel distance in of each frame of the video file using an image processor (In our case MATLAB IMAGE TOOLBOX), depicted in figure 26. After gathering the pointing error pixel distance of each frame, we convert the pixel distance to real time distance. This is accomplished by utilizing 'Elevation B' (refer to figure 20) as a reference measurement to get the full diameter of the laser footprint, since vibration is absent at this elevation. We use this full diameter pixel distance and its equivalent real time distance as a conversion tool from pixel distance to real time distance, including during the presence of vibration, since pixel to real distance ratio remains intact. The systematic procedure for gathering pointing error we have studied so far can be surmised in the sequential diagram, figure 27.

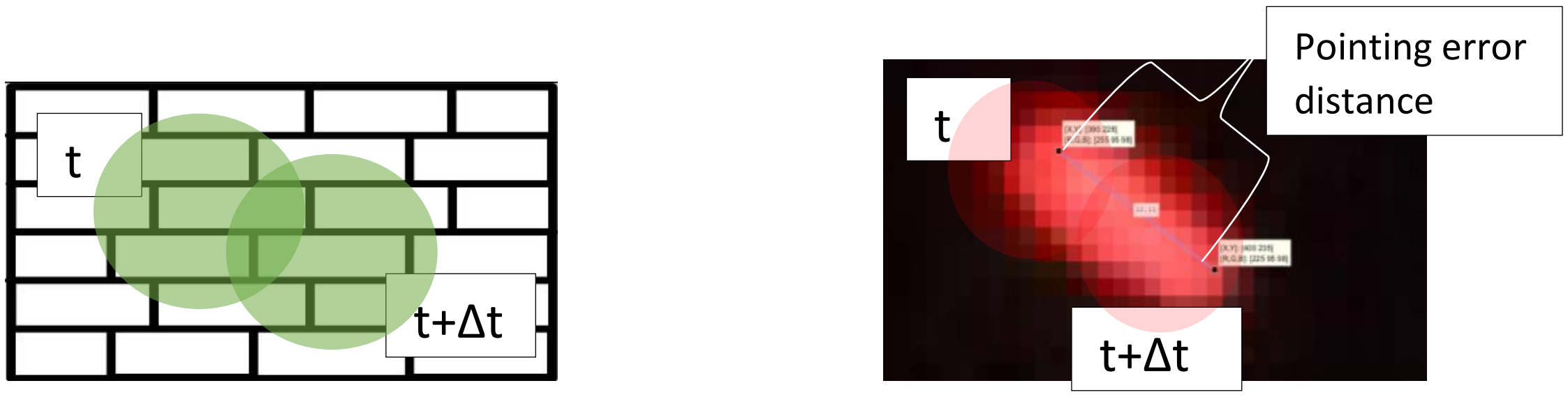

Figure 26. Measuring pointing error distance.

To picture the volume of data we handled for our experiment, we recorded the receiver at 30 fps for 5 sec; hence, 150 frames were candidates for gathering the pointing error. Hence, a total of 150 data sets were used, each data set possessed ‘x’ and ‘y’ components. This data set was used to conceive the data table, Table [2], and to plot the results in figure 18, in section A of ‘Experiment results’.

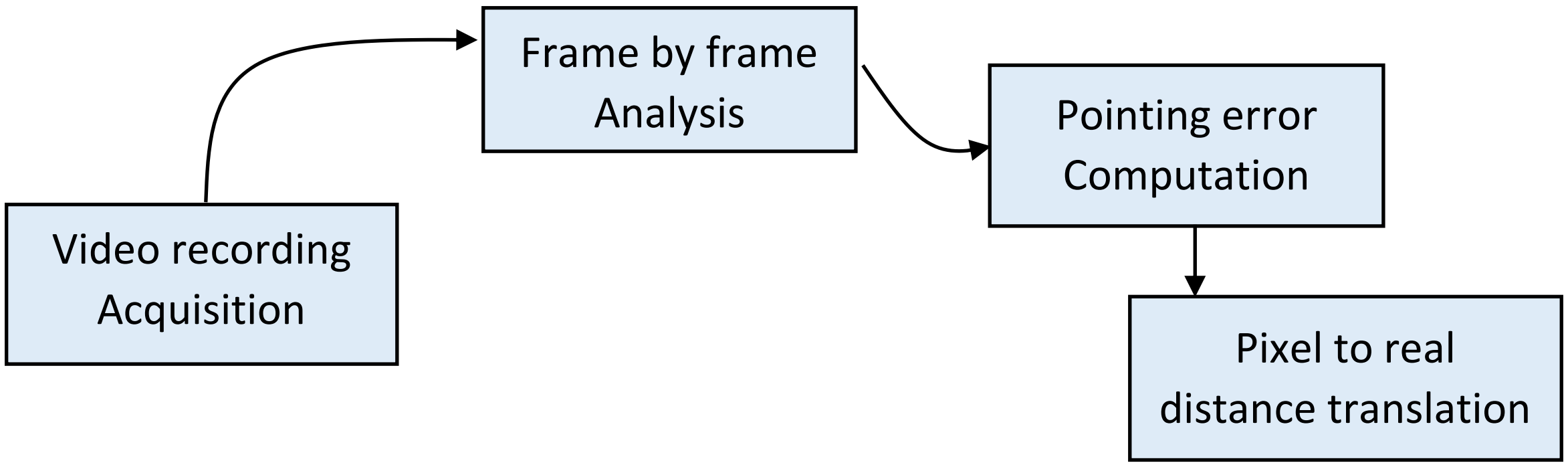

Figure 27. Diagram of the methodology of conducting and analyzing our experiment.

C. Theoretical prediction

In the previous section, we studied a practical technique to measure pointing error. However, in scientific research we are keen in obtaining the measurement merely from scientific formulas and equations using theoretical analysis. To theoretically obtain the pointing error, we should first understand how pointing error originates.

As we have already covered in discussing misalignment error, pointing error of FSO link in free space is predominantly induced by wind or propeller.

Our experiment focuses on propeller-induced vibration. However studying wind-induced vibration is also possible with similar experiment mechanism. Theoretically predicting the vibration of a HAP may not be a very easy task, our research was comprised of electrical engineers and perhaps no mechanical engineers. We sought assistance from Mechanical Engineering professors in KAIST, who specialize in mechanical vibration field such as, Prof. Kim, Kwang-Joon; however, we did not get response. Hence, our approach to solve the vibration problem is with basic vibration theory but accurate physics.

FSOC HAPs floating or sailing in free space in the stratosphere possess more than one degree of freedom. A convenient and popular vibration response solving method for systems with multiple degrees of freedom (MDOF) such as satellites and airplanes is known as modal analysis of the response of MDOF. MDOF of HAPs modal analysis equation is given by the following second order differential equation (4.1). [13]

$$M\ddot{\boldsymbol{x}} + C\ddot{\boldsymbol{x}} + K\boldsymbol{x} = B\boldsymbol{F}(t) \qquad (4.1)$$

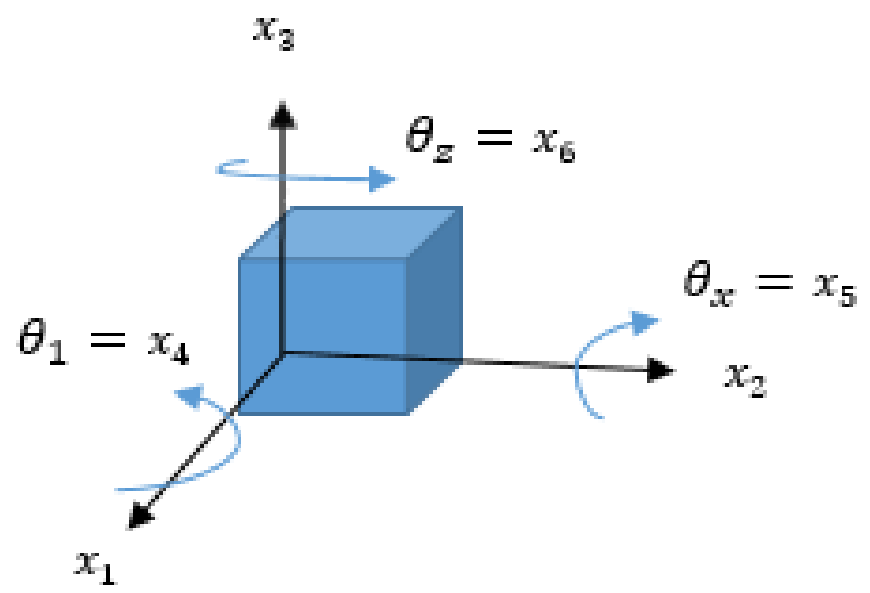

For propeller-induced vibrations, we choose the force provided by the propellers in place of $\boldsymbol{F}(t)$ in equation (4.1), to obtain a polished equation (4.2). Whereas for wind-induced vibrations, we choose the wind force "attacking" the HAP in place of $\boldsymbol{F}(t)$ in equation (4.1) and we arrive at (4.3).

$$M\ddot{\boldsymbol{x}} + C\ddot{\boldsymbol{x}} + K\boldsymbol{x} = B\vec{\boldsymbol{F}}_{\boldsymbol{Propeller}} \qquad (4.2)$$

$$M\ddot{\boldsymbol{x}} + C\ddot{\boldsymbol{x}} + K\boldsymbol{x} = B\vec{\boldsymbol{F}}_{\boldsymbol{Wind}} \qquad (4.3)$$

Where,

M is the HAP mass diagonal matrix, $\boldsymbol{x}$ is the column matrix containing the six degrees of freedom variables, $C$ is the damping coefficient matrix and $K$ is the stiffness matrix.

$$\vec{\boldsymbol{F}}_{\boldsymbol{Wind}} = A\rho_{strat}\vec{\boldsymbol{v}}_{\boldsymbol{wind}} \qquad (4.4)$$

Where, $\rho_{strat} < 0.2$ kg/m$^3$, but varies with temperature**.** $A$ is the area of impact on the HAP by the wind and $b$ is the drag force constant.

$$\vec{\boldsymbol{F}}_{\boldsymbol{Propeller}} = m_{HAP}\vec{\boldsymbol{a}}_{\boldsymbol{HAP}} - b\boldsymbol{v}_{\boldsymbol{HAP}}^{\boldsymbol{2}} \quad (4.5)$$

We now solve the MDOF equation using Google loon's known parameters [14] , by means of Runge–Kutta method in Mathcad prime 5.0.0 using typical values of quadcopter mass (0.7kg), propeller force (lift force, mg), stiffness ($10^3$ N/m and $10^3$ N/mrad) and damping coefficient ($0.002K$). The vibration component solution is plotted in figure 28, in time domain. In figure 28, we can evidently see how the angular vibration effect on pointing error dominates the linear vibration effect with a ratio of nearly 80:1. Subsequently, in figure 29, we plot both experimentally measured value (refer to figure 18) and theoretically predicated value using MDOF. To estimate the theoretical pointing error prediction inaccuracy, we compare the measured and the theoretical pointing error magnitude mean values as shown in table [3], which we obtain around the inaccuracy to be 11.33%.

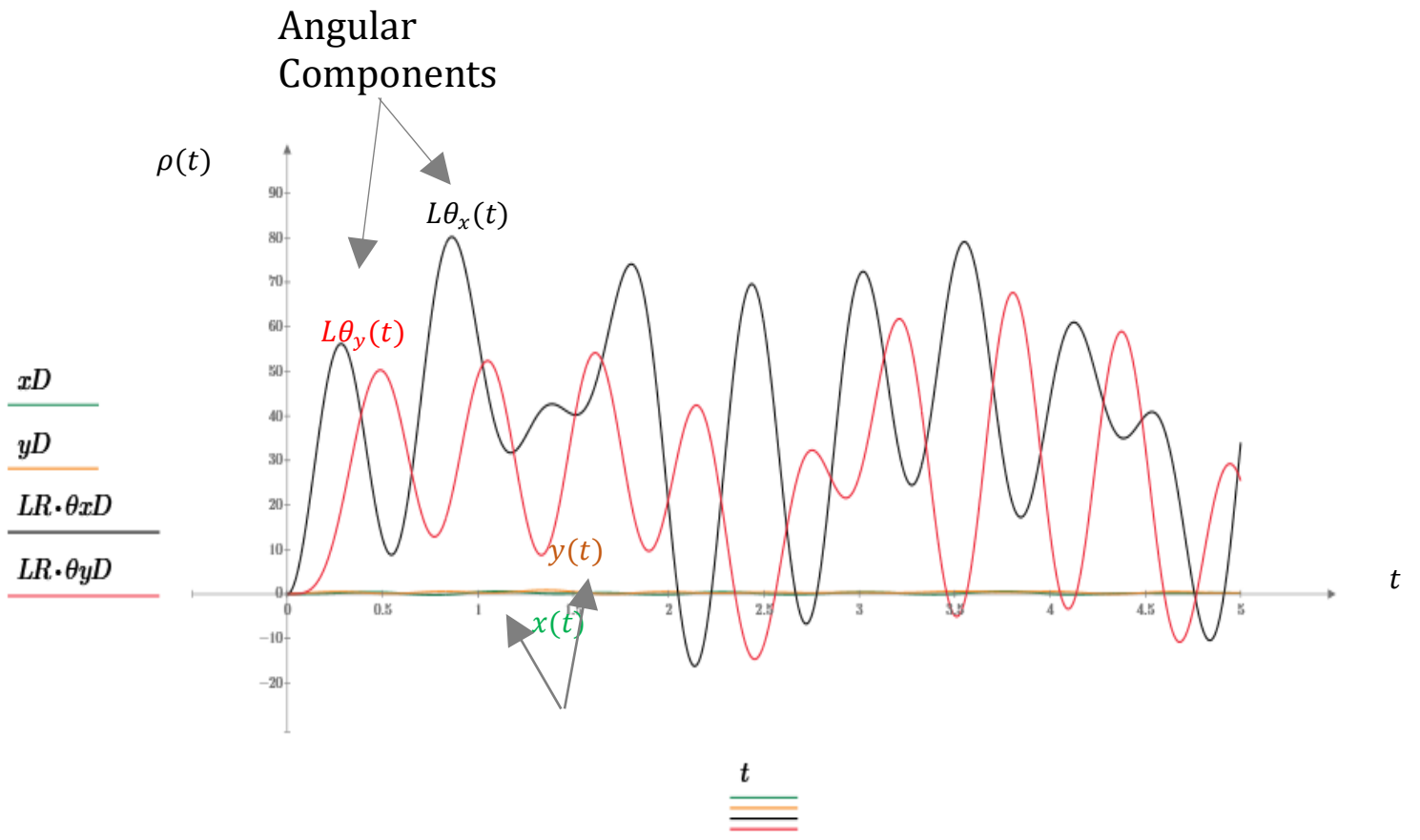

Figure 28. Theoretical prediction plot of the experiment in time domain

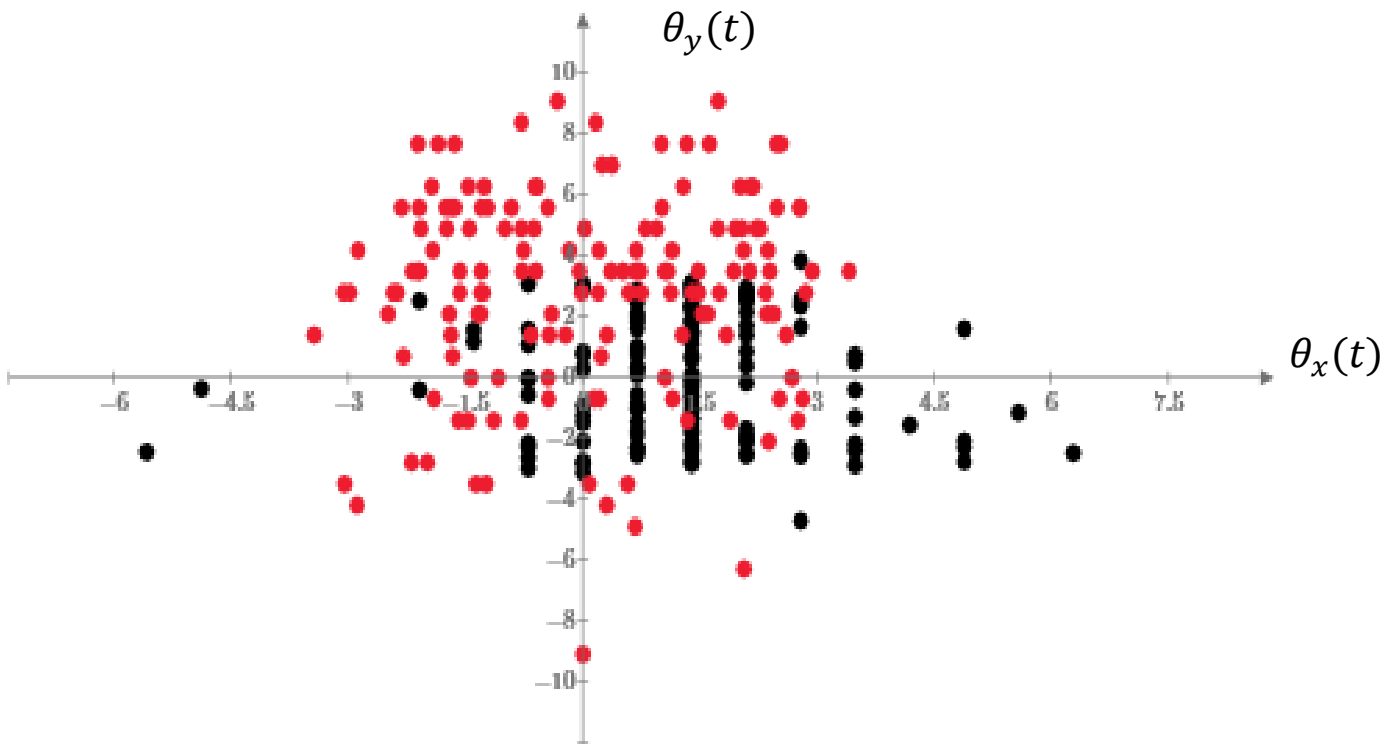

Figure 29. Comparison of theoretical prediction with experiment measurement.

| Statistical Parameter | $\mu_{\theta_x}$ | $\mu_{\theta_y}$ | $\mu_{\lvert\theta\rvert}$ | $\sigma_{\theta_x}$ | $\sigma_{\theta_y}$ | $\sigma_{\lvert\theta\rvert}$ |
|---|---|---|---|---|---|---|
| Measured value (mrad) | 2.8749 | 1.2746 | 6.2563 | 3.2768 | 1.5535 | 12.9700 |
| Theoretical Prediction Value (mrad) | | | | | | 11.500 |
| Prediction error | | | | | | 11.33% |

Table [3]. Error computation for theoretical prediction.

Now that we have obtained a 1 in 10 inaccuracy prediction confidence, we make effort to predict the statistical pointing error distribution of the emerging FSOC project 'Project Loon', as shown in figure 30.

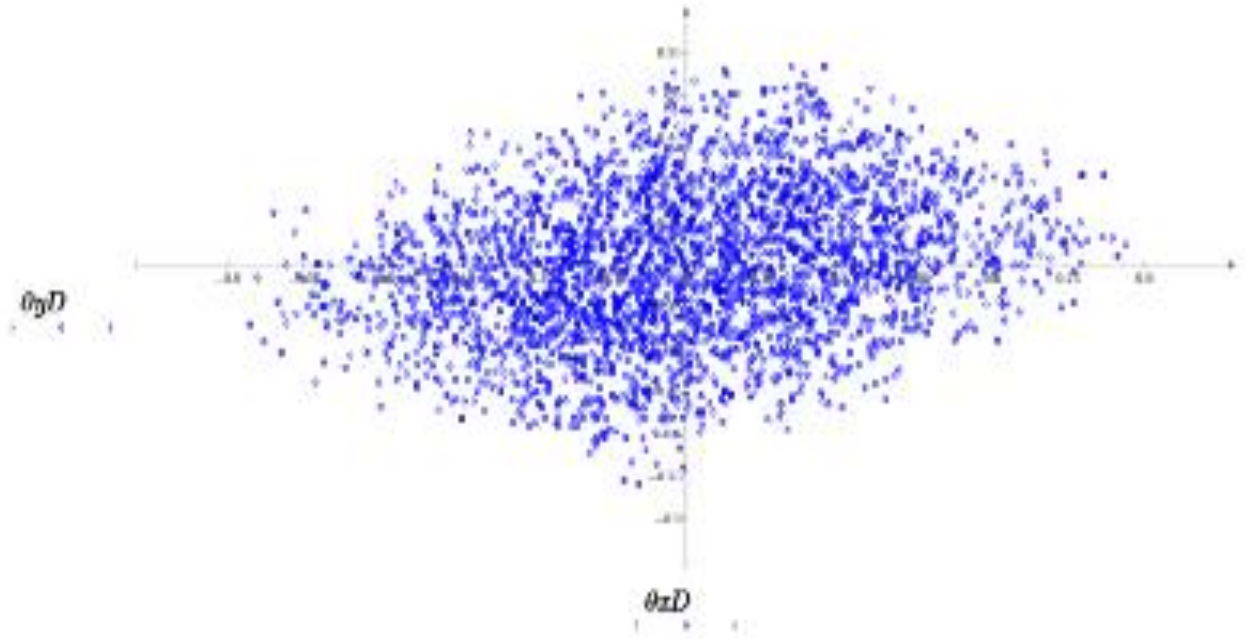

Figure 30. Pointing error prediction for project loon.

Figure 30 is the pointing error plotted for 'typical' conditions (5 m/s resultant wind, no propeller), what happens if the effect of wind is suddenly reduced. Wind induced vibration, with calm wind < 1 m/s, resultant wind would result a more scattered distribution but less angular vibration magnitude, as shown in figure 31; typical example of this wind is during lunch of FSO balloons to space. In contrast, if the speed of the wind is suddenly increased, turbulent wind at 30m/s, clustered distribution results but the angular vibration magnitude is escalated, as depicted in figure 32.

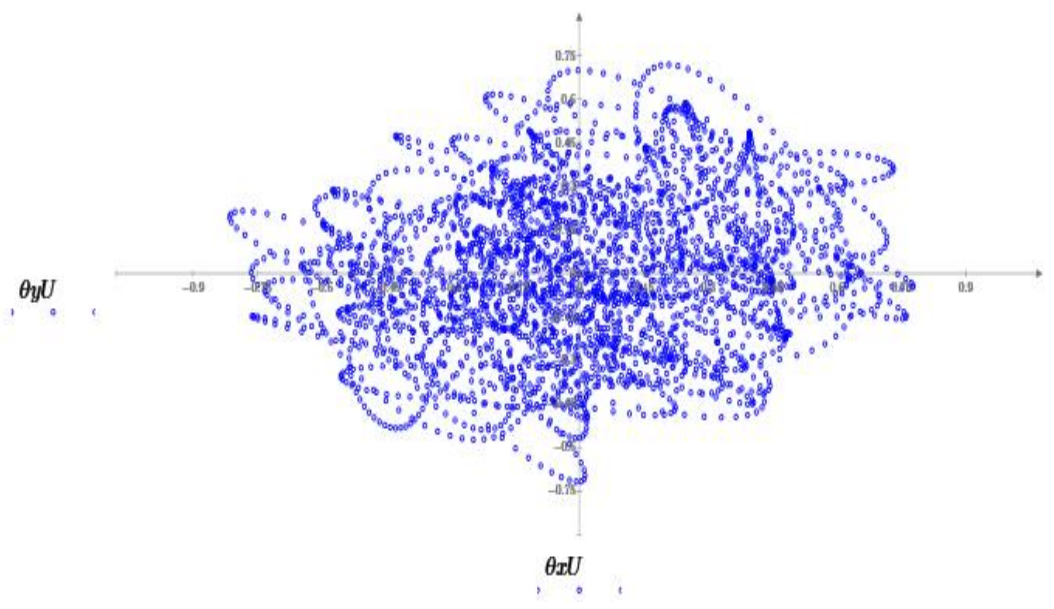

Figure 31. Wind induced vibration, with calm wind < 1m/s resultant wind.

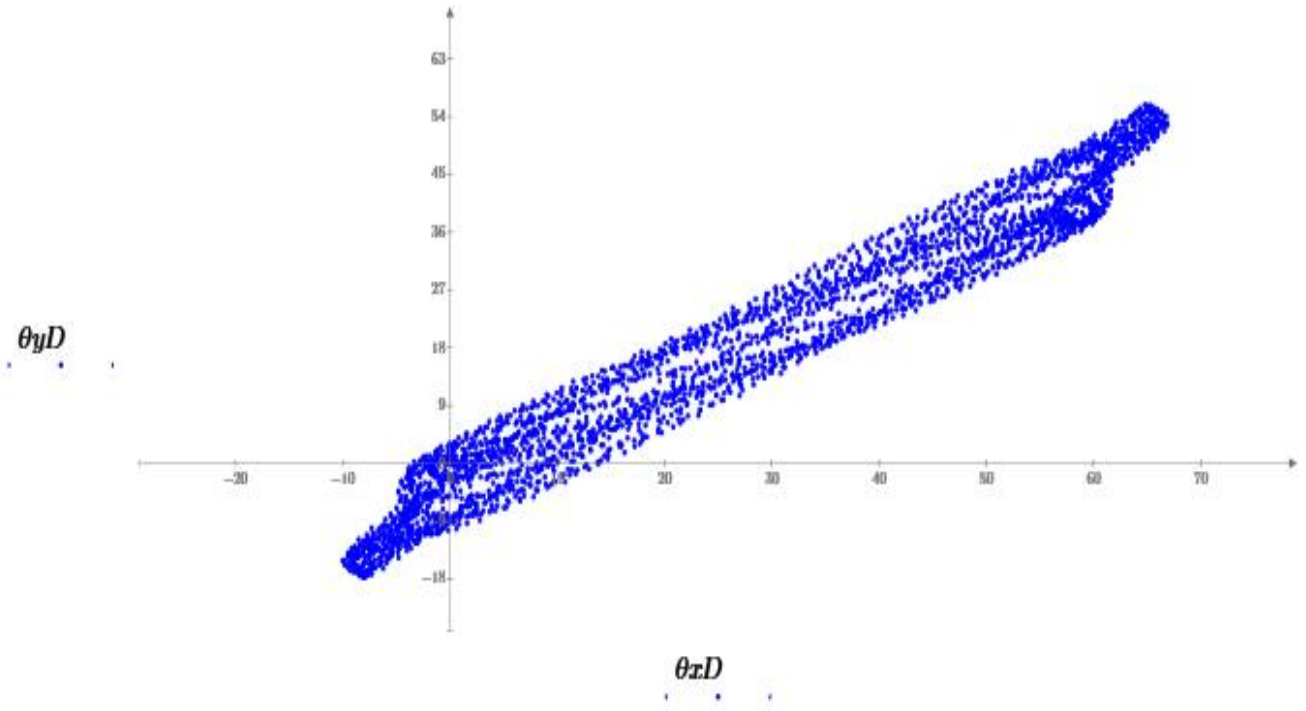

Figure 32. Turbulent wind at 30m/s.

We may also be interested to understand, the wind induced vibration effect in each dimension by discerning the effect of unidirectional wind. The effect of uni-directional wind on wind-induced vibration pointing error distribution is depicted in figure 33. Figure 33 suggests that ‘z’ direction winds are the prominent cause of wind induced-vibration, whereas ‘x’ direction winds affect the least.

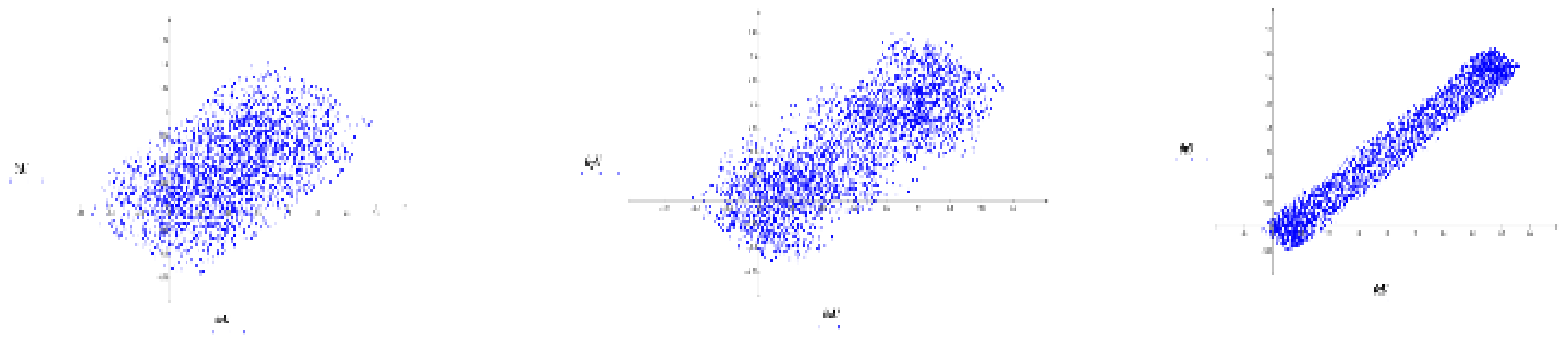

‘x' directional wind ‘y’ directional wind ‘z’ directional wind

10m/s unidirectional wind

Figure 33. Unidirectional wind effect on pointing error distribution.

Based on Figure 33, we can summarize for project loon, the effect of unidirectional wind on misalignment error is presented in table [4].

| Unidirectional wind | Effect on $\tau_{Pointing}$ |
|---|---|
| ‘x' direction | Low |
| ‘y' direction (along gravity path) | Medium |
| ‘z' direction (along FSO link path) | High |

Table [4]. Effect of unidirectional wind on $\tau_{Pointing}$.

# V. Conclusion

In summary, our report primarily introduces FSO and FSOC in stratosphere, followed by a discussion on the causes of attenuation in FSOC channel. In this report, we focus on one of the attenuation factors, misalignment error. We discuss misalignment error and how it results in attenuation. We also discussed the two factors that result in misalignment error, wind-induced vibration and propeller-induced vibration of HAPs. However, due to the lack of sufficient vibration triggered misalignment error research, we arranged an experiment to measure the misalignment error caused by propeller-induced vibration using RC drone. We theoretically predicted the pointing error using MDOF and made comparison with the measured pointing error from the experiment. From the comparison, we concluded that theoretical prediction of pointing error using MDOF results in 11.33% inaccuracy. Finally, our research on pointing error and the experiment methodology could aid researchers in the design of PAT for FSO systems in stratosphere.

Future study

The study of misalignment error will open a new branch to designing systems utilizing PAT. One suggested technique for compensating pointing error is, auto-tracking jets and automated receiver gimbal using CMOS camera. From our research, we can suggest that two orders of pointing error could be resulted and hence, two methods to fix the pointing error. First order refers to small pointing error magnitude and second order refers to large pointing error magnitude. Referring to figure 34, a first order pointing error could be corrected by adjusting the gimbal of the receiver as shown in figure 35, whereas second order pointing error requires larger adjustment of the receiver, hence we can fine-tune the receiver of the HAP using jet propulsion that adjust the HAP to a position of interest.

Another technique is increasing damping of HAPs, increasing the damping of HAPs reduces the magnitude of angular vibration of HAP transmitters by damping their angular shakiness, resulting in reduced pointing error at the receiver. The damping of HAPs could be enhanced by using stiff materials in HAP building material and effective designs that hinder the HAP from recurrent shaking. Less angular shaking of HAPs implies reduced magnitude of pointing error and hence reduced misalignment power loss.

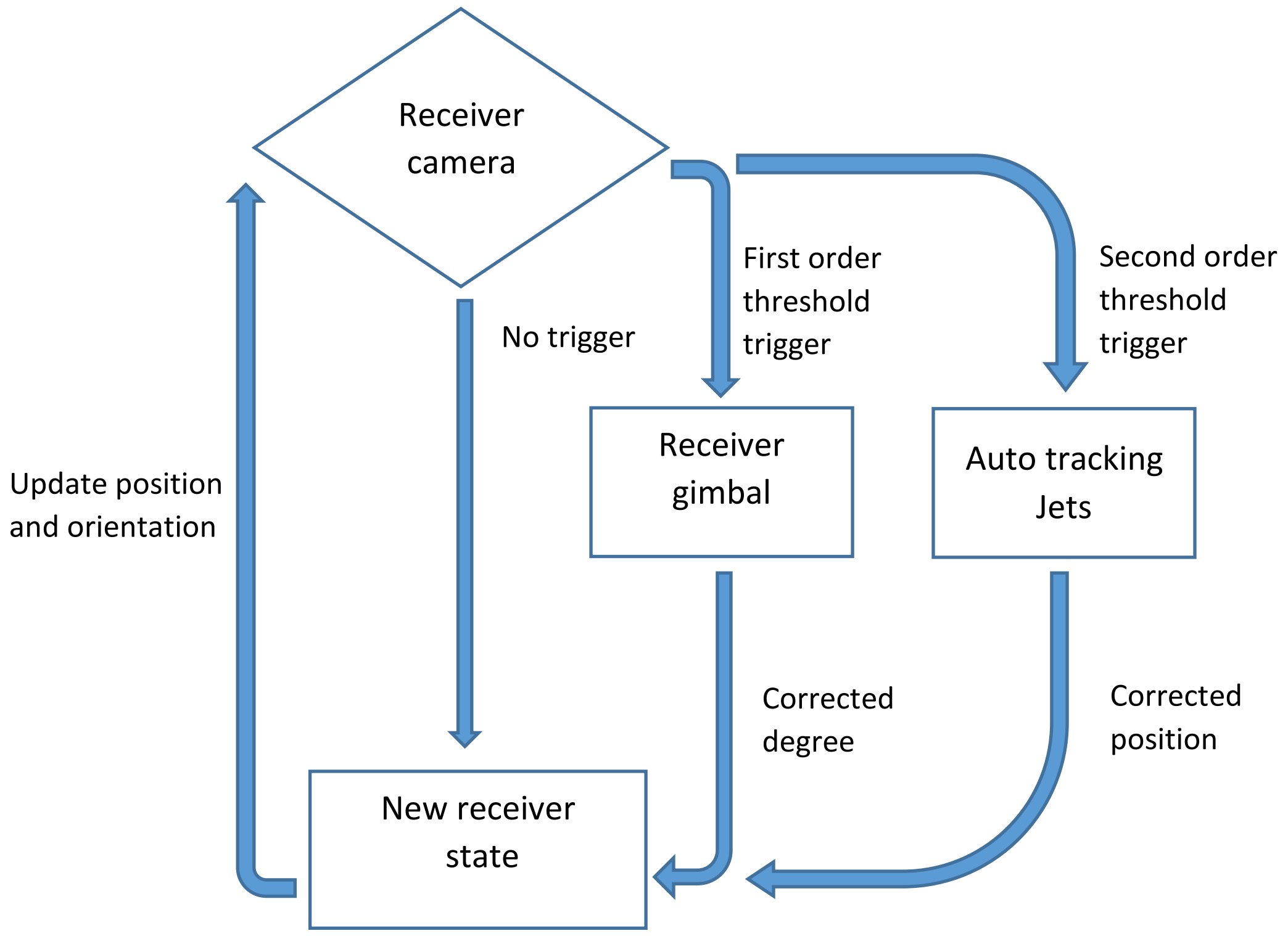

Figure 34. PAT system to reduce power loss.

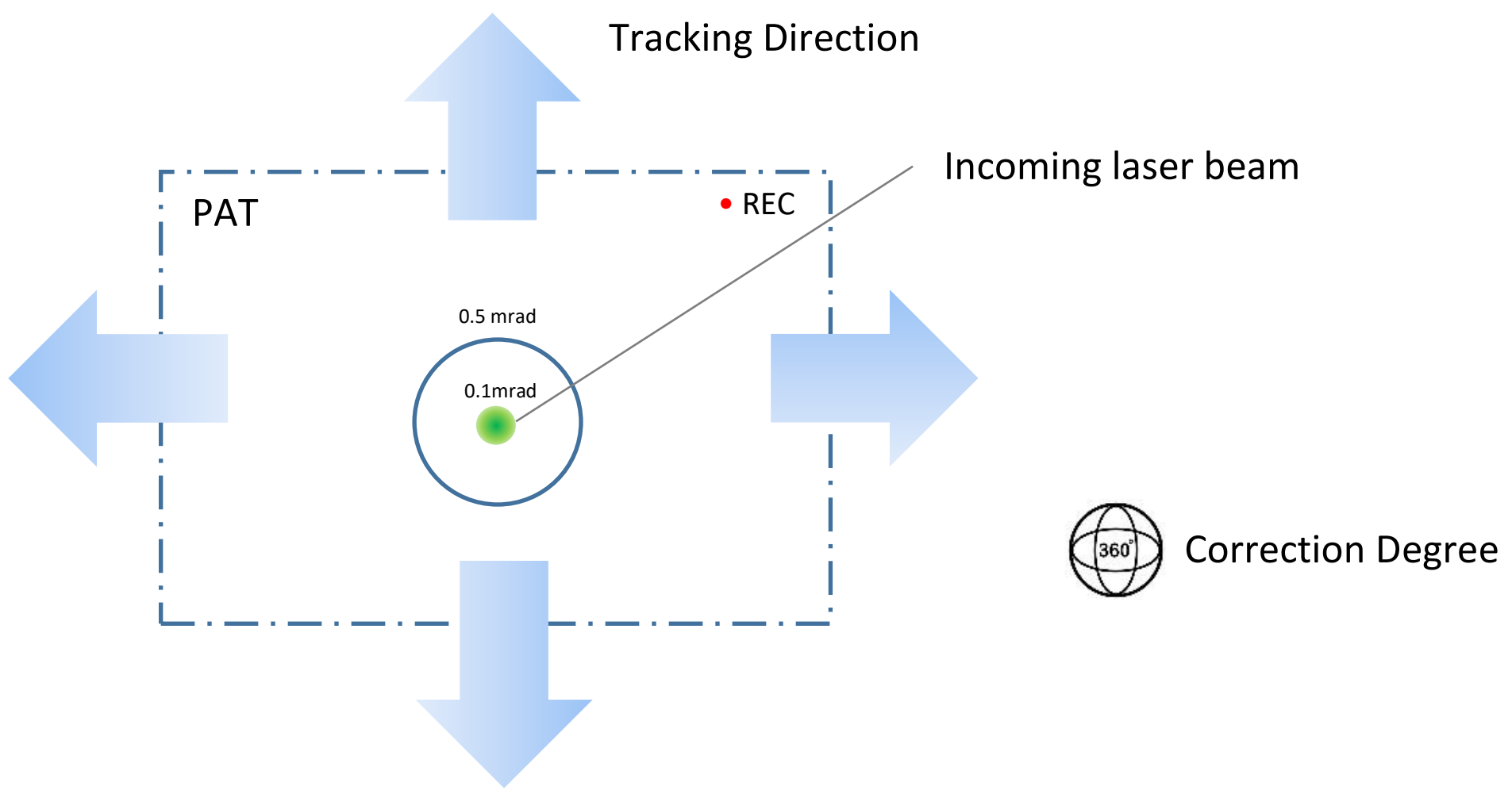

Figure 35. Gimbal correction to align incoming laser beam.

# Appendix A

FSO: Free space Optics

FSOC: Free Space Optical Communication

HAP: High Attitude Platform

UAV: Unmanned Aerial Vehicle

PAT: Pointing Acquisition and Tracking

MDOF: Multiple Degrees of Freedom